\documentclass[10pt,conference]{IEEEtran}
\usepackage{dsfont}
\usepackage{algorithm}
\usepackage{algpseudocode}
\usepackage{listings}
\usepackage{amsmath}
\usepackage{mathtools}
\usepackage{listings}
\usepackage{lstlinebgrd}
\usepackage{tipa}
\usepackage[super]{nth}
\usepackage{subfig}
\usepackage{tabularx}

\usepackage[font={small,bf}]{caption}
\usepackage{listings, multicol, caption}
\usepackage{lstlinebgrd}
\usepackage{lipsum}
\usepackage{multirow}
\usepackage{paralist}
\usepackage{dsfont}
\usepackage{enumitem}
\usepackage{url}

\def\BibTeX{{\rm B\kern-.05em{\sc i\kern-.025em b}\kern-.08emT\kern-.1667em\lower.7ex\hbox{E}\kern-.125emX}}
    
\begin{document}


%
\title{Branch Prediction Is Not A Solved Problem: Measurements,
  Opportunities, and Future Directions}
\author{
  \IEEEauthorblockN{Chit-Kwan Lin and Stephen J. Tarsa}
  \IEEEauthorblockA{\emph{Intel Corporation} \\ Santa Clara, CA}
  \IEEEauthorblockA{\{chit-kwan.lin, stephen.j.tarsa\}@intel.com}

}




%

%
\maketitle
\begin{abstract}
Modern branch predictors predict the vast majority of conditional
branch instructions with near-perfect accuracy, allowing superscalar,
out-of-order processors to maximize speculative efficiency and thus
performance.  However, this impressive overall effectiveness belies a
substantial missed opportunity in single-threaded instructions per
cycle (IPC).  For example, we show that correcting the mispredictions
made by the state-of-the-art TAGE-SC-L branch predictor on SPECint
2017 would improve IPC by margins similar to an advance in process
technology node.

In this work, we measure and characterize these mispredictions.  We
find that they categorically arise from either (1) a small number of
systematically hard-to-predict (H2P) branches; or (2) rare branches
with low dynamic execution counts.  Using data from SPECint 2017 and
additional large code footprint applications, we quantify the
occurrence and IPC impact of these two categories. We then demonstrate
that increasing the resources afforded to existing branch predictors
does not alone address the root causes of most mispredictions.  This
leads us to reexamine basic assumptions in branch prediction and to
propose new research directions that, for example, deploy machine
learning to improve pattern matching for H2Ps, and use on-chip phase
learning to track long-term statistics for rare branches.
\end{abstract}

\section{Introduction}
\label{sec:intro}

Branch prediction is critical to the performance of modern superscalar
processors~\cite{fog2018microarchitecture, amdBP2016, cbp2016} and is
implemented in dedicated branch prediction units (BPUs).  BPUs work by
training statistical models of branch directions observed as
instructions are retired, and then using these models to predict
unresolved directions for subsequent branches as they are fetched.
BPU predictions drive speculative execution, a key technique for
hiding latency in out-of-order CPUs.

\begin{figure}[t!]
  \includegraphics[clip,width=\columnwidth]{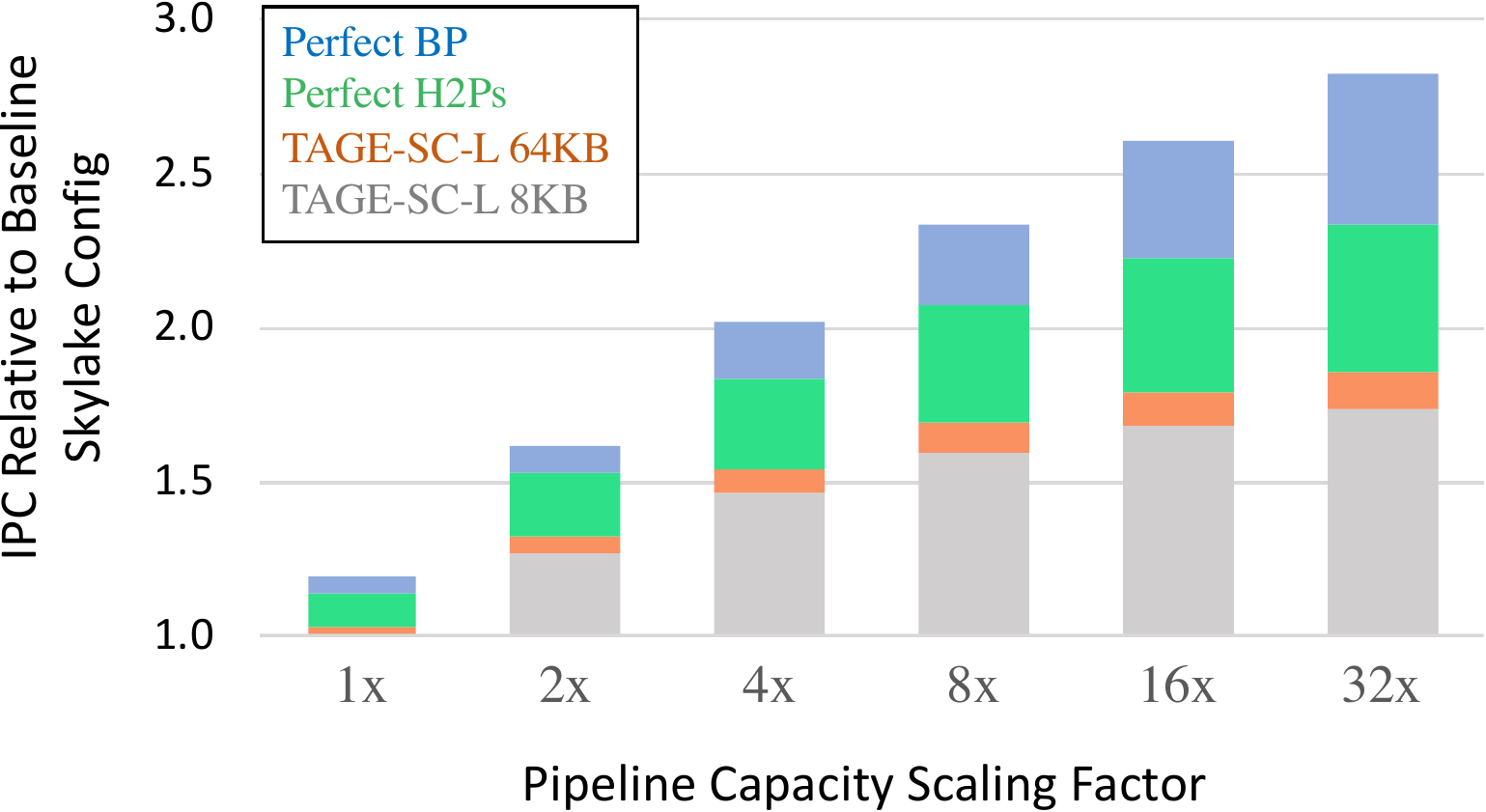}
  \caption{Without better branch prediction, scaling the pipeline
    capacity of an Intel Skylake configuration
    will produce diminishing returns in single-threaded IPC for
    SPECint 2017.}
  \label{fig:ipc-scaling}
\end{figure}

Though state-of-the-art branch predictors achieve near-perfect
prediction accuracy on the vast majority of static branches,
substantial performance gains can be unlocked by correcting their
remaining mispredictions.  Mispredictions delay subsequent
instructions, trigger instruction pipeline flushes, and reduce
speculation efficiency.  For example, Fig.~\ref{fig:ipc-scaling} shows
single-threaded performance for the SPECint 2017 benchmarks on an
execution pipeline based on Intel Skylake, as we simulate future
designs with increased pipeline capacity (i.e., fetch, decode,
execution, load/store buffer, ROB, scheduler, and retire resources) in
the ChampSim simulator~\cite{spec2017, champsim}.  Using the TAGE-SC-L
8KB branch predictor~\cite{seznec_cbp16}, mispredictions represent an
18.5\% instructions per cycle (IPC) opportunity at baseline (1x
scaling).  This gain grows with pipeline scale, e.g., to 55.3\% at 4x
scaling, a magnitude on par with advancing to the next process
technology node.

Microarchitectural advances in branch prediction are thus a source of
large potential IPC gains.  However, we find only marginal
improvements from straightforward scaling of the resources provided to
existing predictors.  Fig.~\ref{fig:ipc-scaling} illustrates this
point: increasing TAGE-SC-L storage eight-fold to 64KB returns just
2.7\% additional IPC in a best-case scenario where no additional
prediction latency is incurred.

In this paper, we perform a deep dive into the causes of these
mispredictions, and demonstrate that fundamentally new approaches to
branch prediction are needed to address them.  We identify two primary
issues: (1) systematically hard-to-predict (H2P) branches; and (2)
rare branches with low dynamic execution counts over tens of millions
of instructions.  In addition to degrading IPC, these branches trigger
inefficient consumption of BPU storage on a large scale, and exhibit
temporal behaviors inconsistent with the short-term statistics
emphasized in known algorithms.  We propose new approaches that deploy
powerful machine learning models to improve the pattern matching that
drives H2P predictions (e.g. using low-precision convolutional neural
networks, as developed in our companion paper~\cite{tarsa2019}), as
well as on-chip phase recognition to capture long term predictive
statistics.

\section{Current State of the Art}
\label{sec:background}

We begin our analysis by reviewing the state of the art in branch
prediction.  Over prior decades, the Championship Branch Prediction
(CBP) challenge~\cite{cbp2016} has served as the primary platform to
compare techniques using common benchmarks and deployment assumptions.
For example, submissions to the most recent CBP held in 2016 were run
on a simulator that (1) standardized the inputs to the BPU to include
the instruction pointer (IP) value, the instruction type, the branch
target, and the observed direction for conditionals; and (2)
restricted BPU storage to 8KB or 64KB, but imposed no restriction on
prediction latency.  These assumptions are compatible with ChampSim,
which we use to close the loop from prediction accuracy to core IPC
for CBP2016 submissions.

Depending on the algorithm, BPUs typically organize raw data into three modalities:
(1) the \emph{global branch history}~\cite{mcfarling93}, 
which is an ordered sequence of recently executed branch directions at any 
point in a program; (2) each branch's \emph{local history}~\cite{yeh92}, which is 
the ordered sequence of directions taken by that branch in the past; and (3) the
\emph{path history}, which consists of the IP values
from recent branches.  CBP2016 submissions model this data using 
the following algorithms:

\medskip
\noindent \textbf{Partial Pattern Matching} (PPM)~\cite{cleary84,
  mudge96} compares a sequence of data against previously observed
sequences of increasing length, and returns the longest exact match. A
PPM branch predictor is implemented by hashing history data over
various lookback windows into tagged table entries that track
directions with a saturating counter.  PPM predictors achieve best
performance when many history lengths are tracked, and both the number
of lengths and the number of table entries used per length are the
primary drivers of their storage/accuracy tradeoff.

\medskip
\noindent \textbf{Perceptron Predictors} mitigate a shortcoming of PPM's
exact pattern matching by learning weights on different history
positions~\cite{jimenez01, jimenez03}.
This improves accuracy when two branches' directions are correlated 
by damping \emph{uncorrelated} history data. 
For PPM, uncorrelated or noisy history data explodes the number of
unique sequences associated with branch statistics; perceptron predictors
more compactly capture correlations by instead training and storing
positional weights. At prediction time, weights are multiplied by a global 
history sequence, summed, and thresholded to generate a prediction. 

\medskip
\noindent \textbf{Domain-Specific Models} fit BPU data to templates of 
program execution behavior. Examples include loop predictors that predict exit
conditions~\cite{sherwood00}, the Wormhole predictor and Inner-Most
Loop Iteration counter (IMLI) that track correlations between branches
in nested loops~\cite{albericio14, seznec2015inner}, and the
Store/Load Predictor, which tracks data dependencies affecting
branch conditions~\cite{farooq13}.  These predictors are derived
from detailed expert analysis, and target 
specific program behaviors found to cause mispredictions in design-time benchmarks.

\medskip
\noindent \textbf{Ensemble Models} generate a single prediction 
from multiple trained models, implementing a form of \emph{boosting}. 
For example, the \emph{statistical corrector}
uses a perceptron-like model to apply weights to predictions of constituent
predictors.\\

\textbf{TAGE-SC-L} is the CBP2016 winner, and we focus on it in this
paper. It implements an ensemble predictor that combines PPM
predictions from histories whose lengths follow a geometric-series
(TA\textbf{GE}) with the IMLI loop predictor (\textbf{L}). The
statistical corrector (\textbf{SC}) arbitrates between available
predictions.

\section{Misprediction Characteristics}
\label{sec:mispred}
Below, we describe the two datasets we used to quantify
mispredictions.

\renewcommand{\arraystretch}{1.2}
\begin{table*}[th!]
  \centering
  \scriptsize
  \begin{tabularx}{\textwidth}{l|r|*{2}{r|}r|r|r|*{4}{r|}r|r}
    \hline
    \hline
    \multirow{3}{*}{\parbox{1.825cm}{\centering SPECint2017 Benchmark}} &
    \multirow{3}{*}{\parbox{0.75cm}{\centering Avg \# Phases}} &
    \multicolumn{2}{c|}{\# Static Branches} &
    \multirow{3}{*}{\parbox{0.8cm}{\centering Avg. Acc.}} &
    \multirow{3}{*}{\parbox{0.8cm}{\centering Avg. Acc. excl. H2Ps}} &
    \multirow{3}{*}{\parbox{0.8cm}{\centering \# App. Inputs}} &
    \multicolumn{2}{c|}{\parbox{2cm}{\centering H2P Appearance Across Inputs}} &
    \multicolumn{2}{c|}{\parbox{1.5cm}{\centering \# Static H2P Branches}} &
    \multirow{3}{*}{\parbox{1.6cm}{\centering Avg. Dyn. Execs per H2P
        per Slice}} &
    \multirow{3}{*}{\parbox{1.25cm}{\centering \% Mispreds due to
        H2Ps per Slice}}\\
    \cline{3-4}\cline{8-11}
    &
    &
    \multirow{2}{*}{\parbox{0.75cm}{\centering Total}} &
    \multirow{2}{*}{\parbox{0.80cm}{\centering Median per Slice}} &
    &
    &
    &
    \multirow{2}{*}{\parbox{0.80cm}{\centering Total}} &
    \multirow{2}{*}{\parbox{0.8cm}{\centering 3+ Inputs}} &
    \multirow{2}{*}{\parbox{0.80cm}{\centering Avg per Input}} &
    \multirow{2}{*}{\parbox{0.80cm}{\centering Avg per Slice}} & &
    \\
    & & & & & & & & & & & & \\
    & & & & & & & & & & & & \\
    \hline\hline
    600.perlbench\_s & 6.5 & 13,865 & 1,863 & 0.987 & 0.989 &  4 &  62 & 16 & 21.5 &  1 &   93,815  & 17.3\% \\
    605.mcf\_s       & 11.4 & 1,755 &    99 & 0.921 & 0.998 &  8 &  29 & 20 & 19.0 & 10 &  249,195  & 96.9\% \\
    620.omnetpp\_s   & 11.8 & 7,099 &   823 & 0.975 & 0.994 &  5 &  46 & 28 & 28.0 &  8 &   74,630  & 77.6\% \\
    623.xalancbmk\_s &  7.5 & 8,563 & 3,103 & 0.997 & 0.998 &  4 &  28 &  8 & 14.5 &  6 &   75,329  & 28.6\% \\
    625.x264\_s      & 13.9 & 4,892 & 1,068 & 0.946 & 0.975 & 14 &  23 &  7 &  6.0 &  1 &   65,593  & 54.2\% \\
    631.deepsjeng\_s &  9.4 & 3,162 &   856 & 0.946 & 0.963 & 12 &  68 & 49 & 40.0 & 13 &   44,412  & 31.2\% \\
    641.leela\_s     &  8.8 & 3,623 &   582 & 0.880 & 0.960 & 10 &  77 & 68 & 56.5 & 34 &   35,614  & 66.4\% \\
    648.exchange2\_s &  8.4 & 3,765 & 1,330 & 0.986 & 0.992 &  5 &  38 & 19 & 20.0 &  7 &  142,320  & 44.7\% \\
    657.xz\_s        &  7.6 & 2,373 &   211 & 0.897 & 0.980 &  5 & 163 & 50 & 63.0 & 10 &   75,759  & 80.5\% \\
    \hline
    MEAN             &  9.5 & 5,455 & 1,104 & 0.952 & 0.984 &  7 &  59 & 29 & 30.0 & 10 &   95,185  & 55.3\% \\
    \hline\hline
  \end{tabularx}
  \caption{Summary statistics of our SPECint 2017 data set, which
    includes an expanded collection of inputs for each benchmark.  Metrics
    are averaged over 10B-instruction traces from each
    input. Accuracy and H2P statistics are reported for TAGE-SC-L 8KB.
  }
  \label{tab:spec2017}
 \end{table*}


\subsection{H2P branches in SPECint 2017}
We primarily used traces of SPECint 2017 benchmarks to study H2Ps.
This dataset was constructed by first compiling each SPECint 2017
benchmark in the single-threaded ``SPECspeed'' configuration, and then
tracing the resulting binaries over multiple application inputs (i.e.,
``workloads'').  Similar to Amaral \emph{et
  al.}~\cite{amaral2018alberta}, we expanded the set of application
inputs (see Table~\ref{tab:spec2017}) for each benchmark in order to
capture greater diversity in program statistics and invariant
behaviors across distinct application inputs.  Each workload was
traced for 10B instructions, which we post-processed into
30M-instruction slices; this slice length matches the default
granularity of SimPoint phase labeling and maintains consistency with
prior analyses~\cite{john2018}.  The slices were then clustered via
SimPoint~\cite{sherwood02} and labeled accordingly.  Doing so verifies
that our 10B-instruction trace length captures a variety of distinct
application phases (9.5 phases, on average).  However, we emphasize
that the branch misprediction statistics shown in
Table~\ref{tab:spec2017} are \emph{not} collected from just the
SimPoints, but across all 30M-instruction slices of each workload
trace (i.e., 333 slices total for each 10B-instruction workload
trace).  This methodology helps capture stable statistics over time,
i.e., over multiple occurrences of the phases.

Within each 30M-instruction slice of every workload, we screen
branches and identify as H2Ps those that (1) have less than 99\%
prediction accuracy under TAGE-SC-L 8KB, (2) execute at least 15,000
times, and (3) generate at least 1,000 mispredictions in the slice.  Our
screening criteria are chosen to identify branches that exhibit
behaviors consistent with systematic misprediction, and that produce
sufficient history data to train machine learning
models~\cite{tarsa2019}.  We screen using TAGE-SC-L 8KB because it
represents a practical implementation of the state-of-the-art under
common CPU resource budgets that we encounter today; in
Section~\ref{sec:scaling}, we present a limit study of scaling up
TAGE-SC-L resources and its effect on H2Ps.

\begin{figure}[t!]
    \includegraphics[clip,width=\columnwidth]{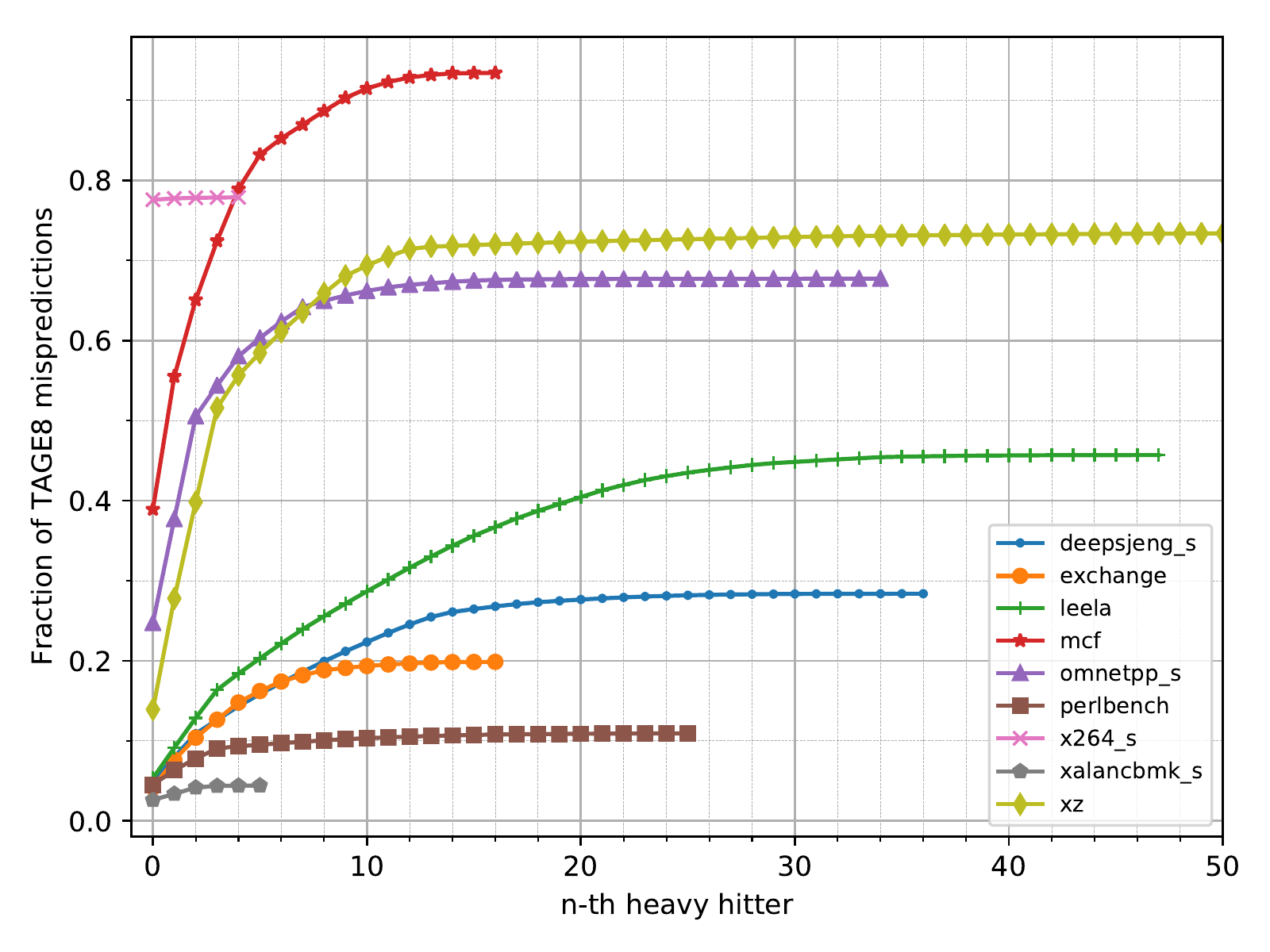}
  \caption{Cumulative fraction of mispredictions due to H2Ps
    for SPECint 2017 benchmarks.  The top five ``heavy hitters"
    account for 37\% of dynamic
    mispredictions on average.}
  \label{fig:hh_ecdf}
\end{figure}

The branch statistics shown in Table~\ref{tab:spec2017} convey two
important messages.  First, there exist a small number of H2Ps that
consistently produce mispredictions every time the application
executes---on average, 29 H2Ps appear in three or more workloads per
benchmark.  These H2Ps present a clear target for specialized
prediction mechanisms.  Second, over all workloads, 55.3\% of the
mispredictions in each 30M-instruction slice are caused by just 10
H2Ps on average. Fig.~\ref{fig:hh_ecdf} plots the cumulative fraction
of mispredictions for H2Ps in each benchmark ranked by total number of
dynamic executions.  We see that the top five in their respective
benchmarks account for 37\% of dynamic mispredictions on average, and
dub these ``heavy-hitters.''
Taken together, this means that \emph{just a handful of
  static branches cause a disproportionately large number of dynamic
  mispredictions}, and that devoting resources to improve their
prediction accuracy is an attractive strategy for increasing
performance.

\begin{table}[t!]
  \begin{tabularx}{\columnwidth}{l|r|r|r|c}
    \hline
    \hline
    \multirow{3}{*}{\parbox{1.5cm}{\centering Application}} &
    \multirow{3}{*}{\parbox{1cm}{\centering Static Branch IPs}} &
    \multirow{3}{*}{\parbox{1.7cm}{\centering Avg. Dyn. Execs per
        Static Branch}} &
    \multirow{3}{*}{\parbox{1.25cm}{\centering Avg. Acc. per Static Branch}} &
    \multirow{3}{*}{\parbox{1cm}{\centering H2Ps}} \\
    & & & & \\
    & & & & \\
    \hline
    \hline
    602.gcc\_s & 6,152 & 715.6 & 0.88 & 5 \\
    Game & 45,996 & 55.2 & 0.73 & 1 \\
    RDBMS & 16,096 & 314.3 & 0.92 & 8 \\
    NoSQL Database & 7,449 & 331.0 & 0.93 & 2 \\
    \multirow{2}{*}{\parbox{1.5cm}{Real-time Analytics}} & 5,595 & 856.0
    & 0.83 & 6 \\
    & & & & \\
    Streaming Server & 3,144 & 1404.7 & 0.78 & 6 \\
    \hline
    MEAN & 14,072 & 612.8 & 0.85 & 5.2 \\    
    \hline\hline
    
  \end{tabularx}
  \caption{Summary branch statistics from large code footprint
    applications under TAGE-SC-L 8KB.  Metrics shown are over 30M-instruction traces.}
  \label{tab:lcf_stats}
\end{table}

\subsection{Rare branches in large code footprint (LCF) traces}
\label{sec:rare_branches}
Table~\ref{tab:spec2017} excludes 603.gcc\_s because we notice that
its much larger code footprint includes many more branches that make 
small but significant contributions to overall mispredictions. To
study this effect further, we collected a set of similarly large code
footprint (LCF) traces.

LCF binaries all have many more static branch IPs per 30M-instruction slice 
than our SPECint 2017 dataset, as shown by comparing Table~\ref{tab:spec2017} to
Table~\ref{tab:lcf_stats}.  We chose applications with static branch counts 
above that of SPEC 2017 623.xalancbmk\_s, the next largest footprint after
603.gcc\_s.  In addition to 603.gcc\_s, we include five additional
applications (a game, a RDMBS, a NoSQL database, a real-time analytics
engine, and a streaming server) which were traced from live
deployments.  For these, we analyze a single 30M-instruction
trace for each application. Though less comprehensive than the
SPEC2017 dataset, this is nonetheless sufficient to illustrate the rare
branch problem in large code footprints.

\begin{figure*}[t!]
    \includegraphics[clip,width=\textwidth]{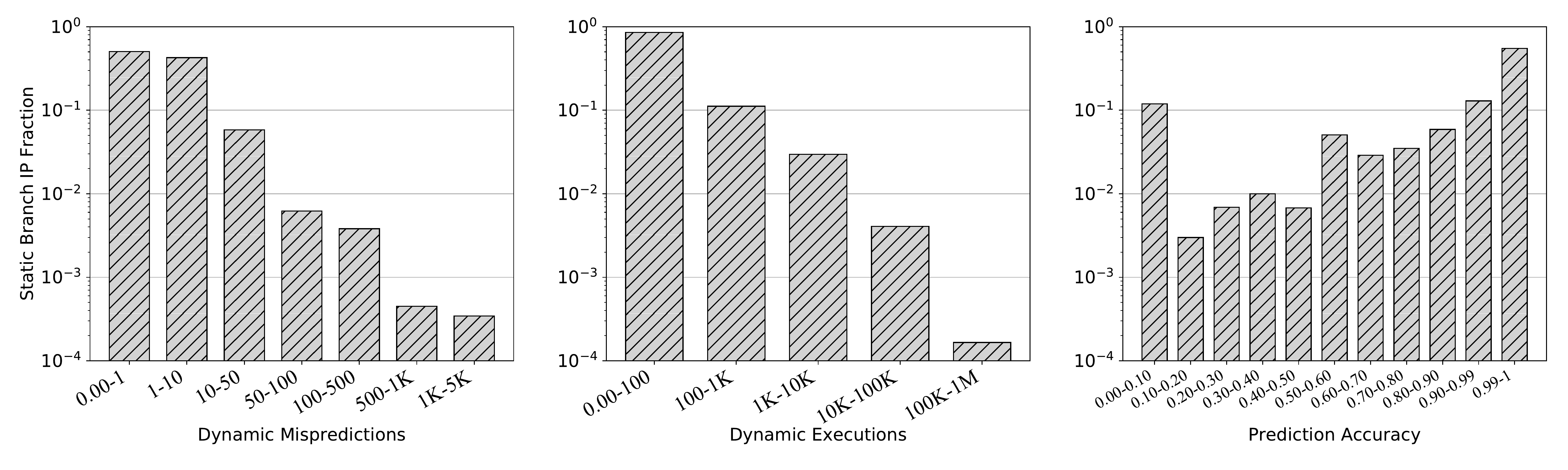}
  \caption{The distributions of dynamic mispredictions (left), dynamic
    executions (middle), and prediction accuracy (right) of branches
    in the LCF data set, under TAGE-SC-L 8KB.}
  \label{fig:hist_lcf}
\end{figure*}

\begin{figure*}[t!]
  \subfloat[]{
    \includegraphics[clip, width=\columnwidth]{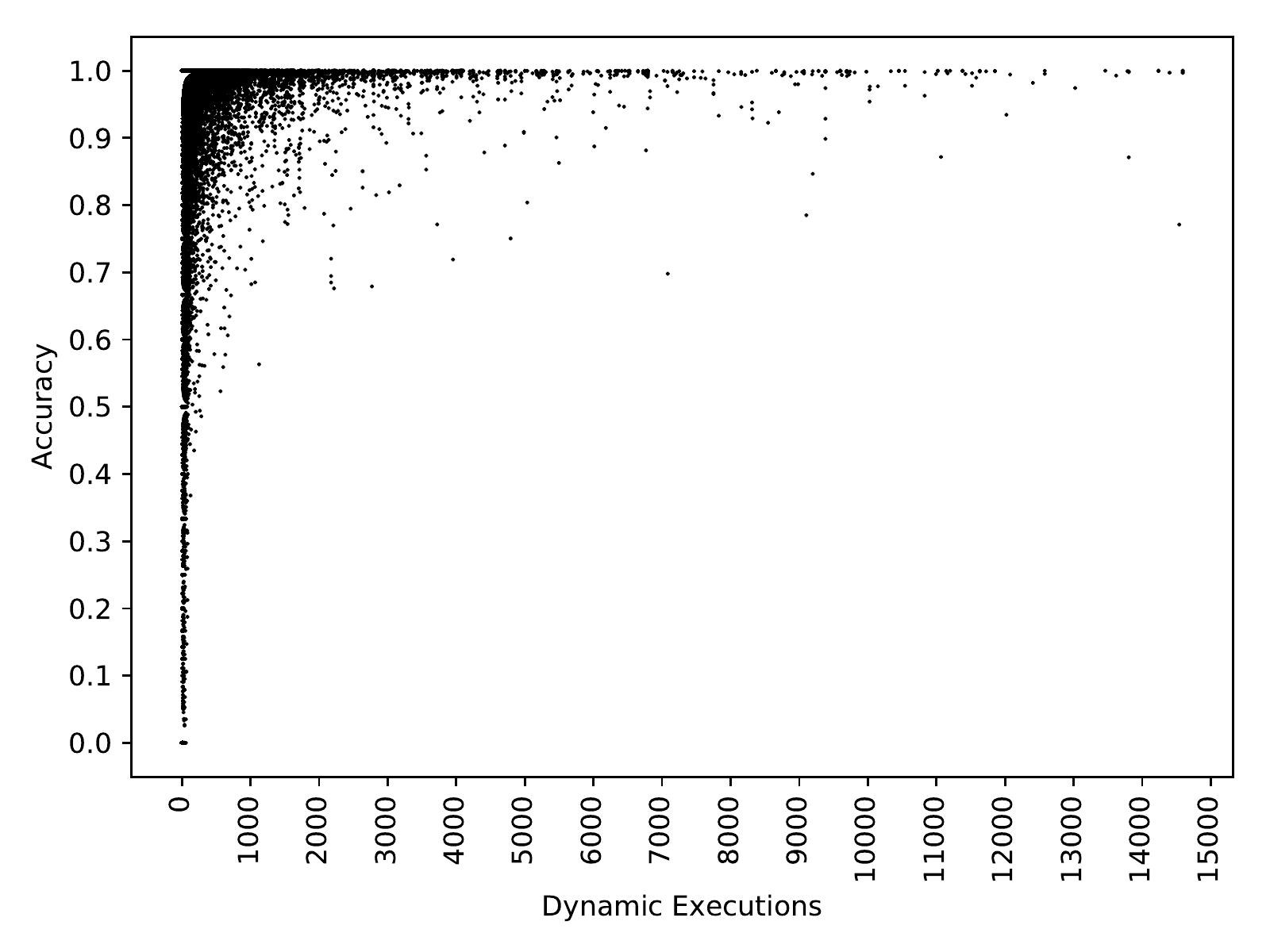}
    \label{fig:scatter_acc_dynexecs}
  }\hfill
  \subfloat[]{
    \includegraphics[clip, width=\columnwidth]{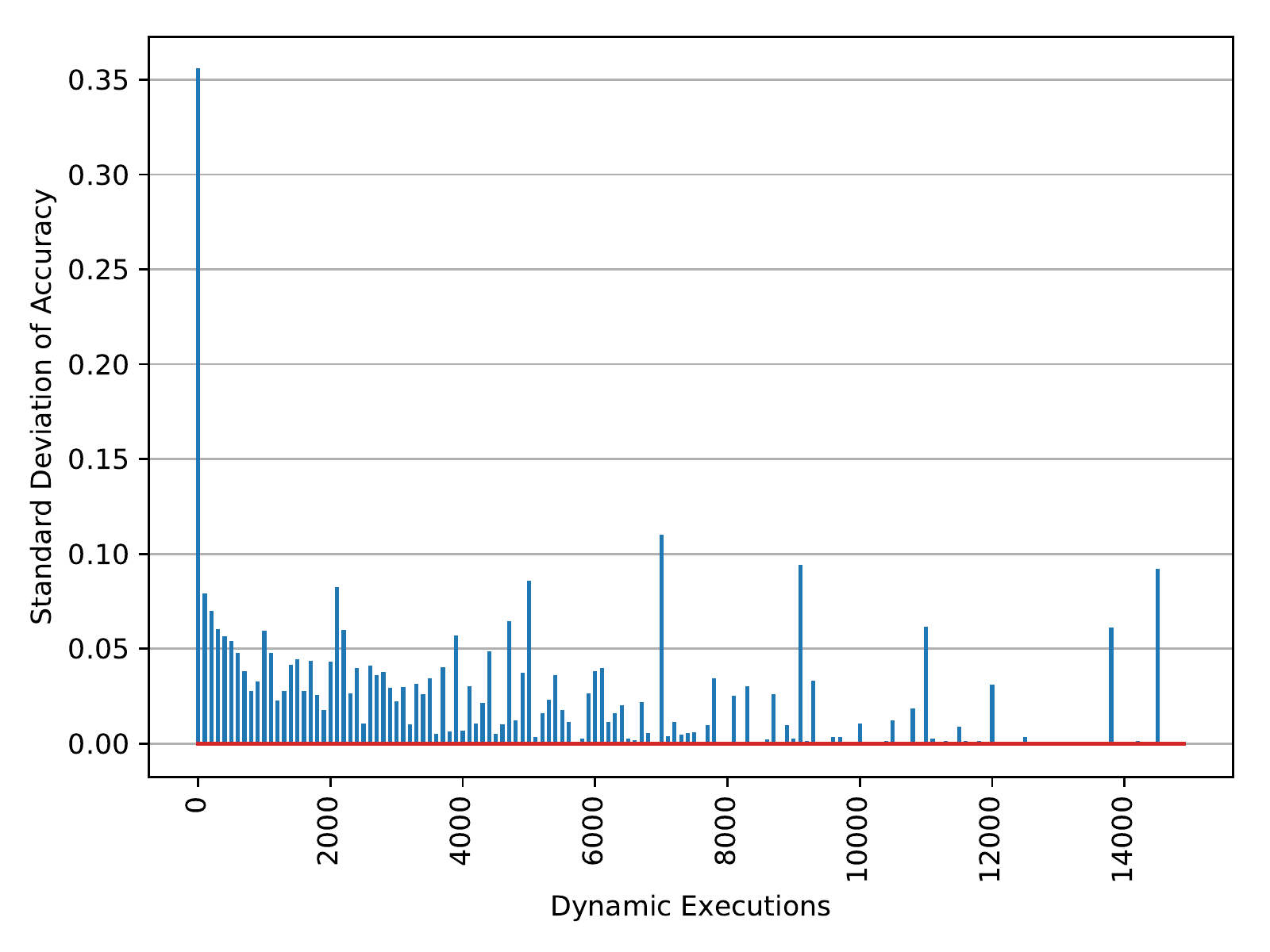}
    \label{fig:binnedstats_acc_dynexecs}
  }
  \caption{\protect\subref{fig:scatter_acc_dynexecs} For the LCF
    dataset, branches with low dynamic execution count have a wide
    spread in prediction accuracy under TAGE-SC-L 8KB.  Each data
    point is a branch. \protect\subref{fig:binnedstats_acc_dynexecs}
    Standard deviation of branches binned by dynamic execution count.
   }
  \label{fig:lcf_acc_vs_dynexecs}
\end{figure*}

Table~\ref{tab:lcf_stats} summarizes branch statistics for the LCF
applications.  There are two observations of note: (1) the average
accuracy of TAGE-SC-L 8KB for these applications is significantly
lower (0.85) than for the SPECint 2017 dataset (0.952); (2) for the
large number of static branches in each application (mean: 14,072),
the average number of dynamic executions per static branch is small
(mean: 612.8).  Fig.~\ref{fig:hist_lcf} further breaks out these
summary statistics into distributions of branches over the entire
dataset.  As we can see, the distribution of dynamic executions
(middle) skews towards the left, with fully 85\% of static branch IPs
executing less than 100 times.  Additionally, the distribution of
dynamic mispredictions (left) skews towards zero, i.e., the vast
majority of branches are predicted with high accuracy.  This is
corroborated by the distribution of prediction accuracy (right), where
55\% of branches are predicted with 0.99 accuracy or greater. Yet,
there is a significant fraction (12\%) of static branch IPs which are
predicted with an accuracy of 0.10 or lower.

Fig.~\ref{fig:scatter_acc_dynexecs} plots dynamic execution count
against prediction accuracy for each static branch IP.  This shows
that \emph{rare branches}, i.e., those with low dynamic execution
counts, have a wide spread in prediction accuracy.  In a sense, this
is unsurprising because rare branches have fewer samples of path and
direction history and thus their collected statistics are
lower-confidence.  Fig.~\ref{fig:binnedstats_acc_dynexecs} quantifies
this spread, showing the standard deviation in prediction accuracy
when we bin dynamic executions (bin width = 100).  The first bin, with
less than 100 dynamic executions, has a standard deviation in accuracy
of 0.35, but this drops off precipitously to just 0.08 for branches
with 100--200 dynamic executions.  

In summary, LCF applications have many rare branches, which are static
branches that are predicted poorly, but execute only a handful of times and thus do not meet 
the H2P criteria as a source of systematic misprediction. 
We also note that both scenarios are present in the above
datasets, but to varying degrees---the SPECint 2017 dataset
showcases H2Ps more than rare branches, whereas the opposite is true
for the LCF dataset.  


\subsection{The effect of CPU pipeline scaling on mispredictions}
\label{sec:pipeline_effect_on_mispreds}
\begin{figure}[t!]
    \includegraphics[clip,width=\columnwidth]{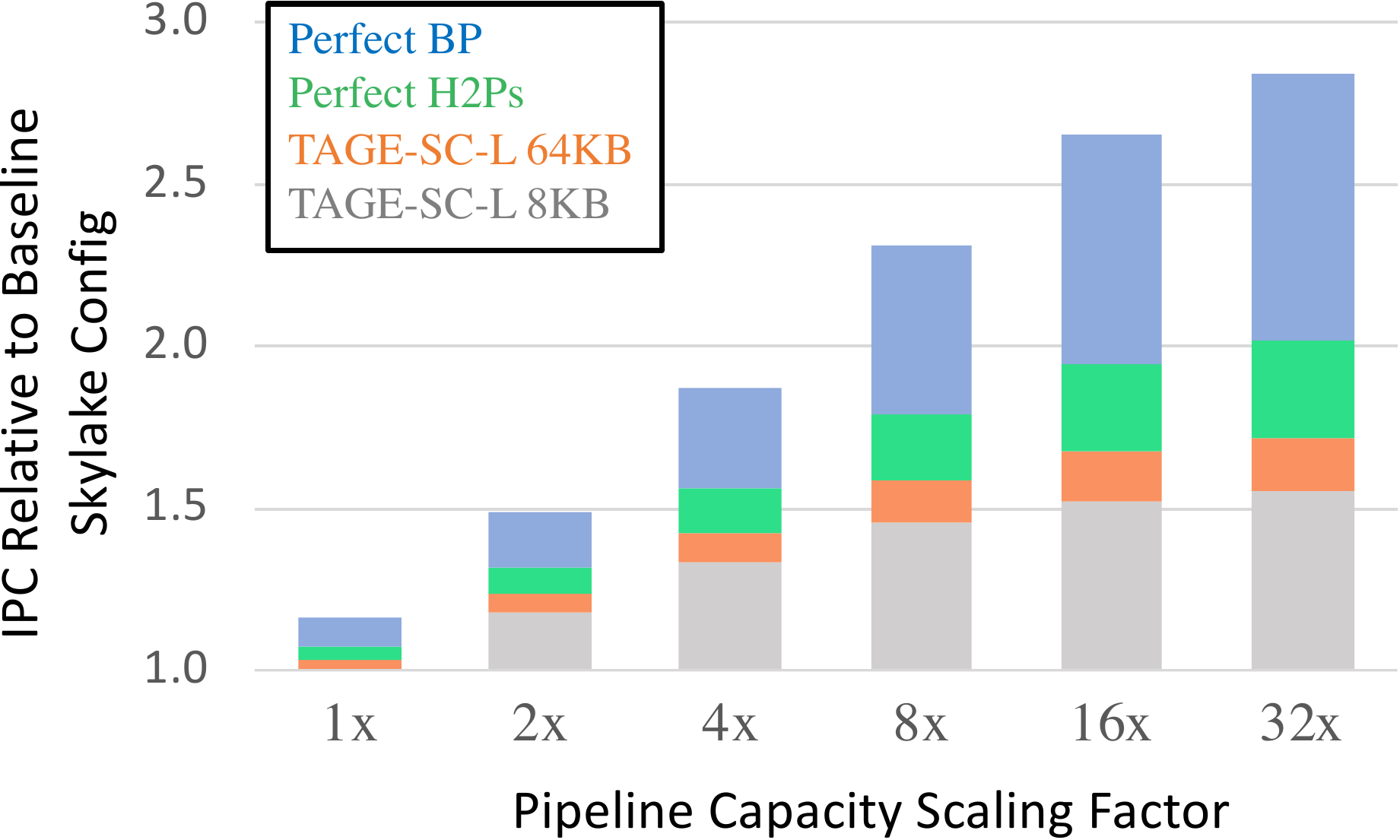}
  \caption{For the large code footprint traces, H2Ps play a
    dramatically diminished role as CPU pipeline scales up. }
  \label{fig:lcf_ipc_scaling}
\end{figure}

One key observation is that H2Ps and rare branches do not have equal
impact on performance when the CPU pipeline is scaled up, e.g., in
future cores.  Fig.~\ref{fig:ipc-scaling} shows that, for the smaller
code footprint applications in SPECint 2017, H2Ps account for 75.7\%
of the potential IPC gain of perfect branch prediction at
baseline. When the pipeline is scaled up, the proportion shifts to
near parity, with H2Ps accounting for 54.8\% of the opportunity.  For
the LCF dataset, non-H2P rare branches play the central role, as shown
in Fig.~\ref{fig:lcf_ipc_scaling}.  At the baseline 1x pipeline scale,
H2Ps represent just 37.8\% of the performance opportunity, and this
even drops slightly to 33.7\% at 32x pipeline scale. Together, this
data shows that \emph{while H2Ps represent an outsized portion of the
  IPC impact for SPECint 2017 on existing CPUs, as pipelines scale and
  application sizes grow, rare branches become an equally important
  source of potential performance.}

\section{Scaling BPU Resources is Not Enough}
\label{sec:scaling}

Aside from the practical limitations imposed by area and latency
constraints~\cite{jimenez2000impact}, simply increasing BPU resources
is insufficient to capture the sizable remaining IPC opportunity due
to branch mispredictions.  As we saw earlier, for a given CPU pipeline
width and depth, scaling TAGE-SC-L global history table sizes, e.g.,
from 8KB to 64KB, as in Fig.~\ref{fig:ipc-scaling}, gives poor
returns.  We next analyze the reasons for this behavior.

\begin{figure*}[ht!]
  \centering
  \subfloat[605.mcf\_s]{
    \includegraphics[height=4cm]{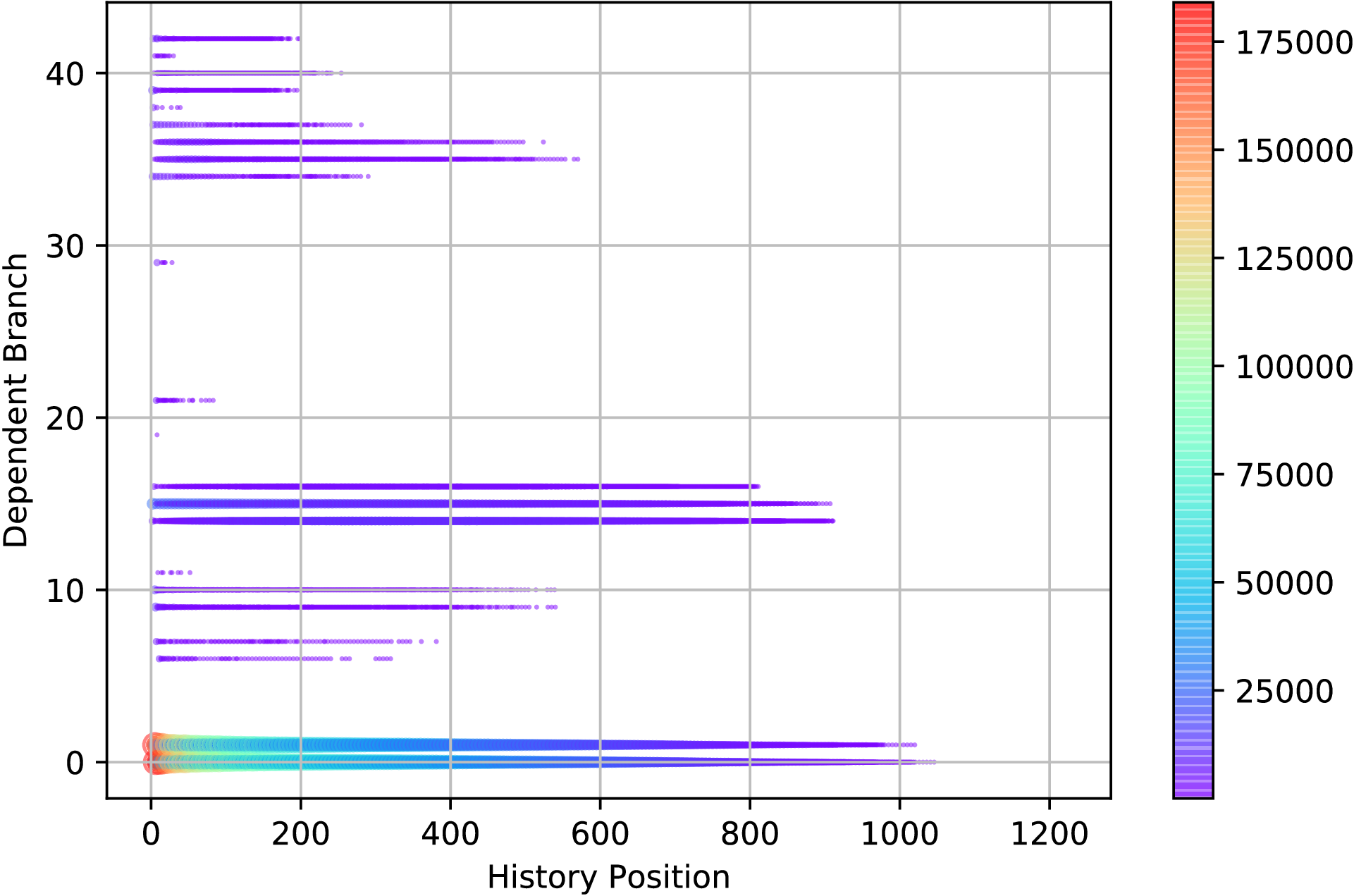}
    \label{fig:depbranch-mcf}
  }
  \subfloat[620.omnetpp\_s]{
    \includegraphics[height=4cm]{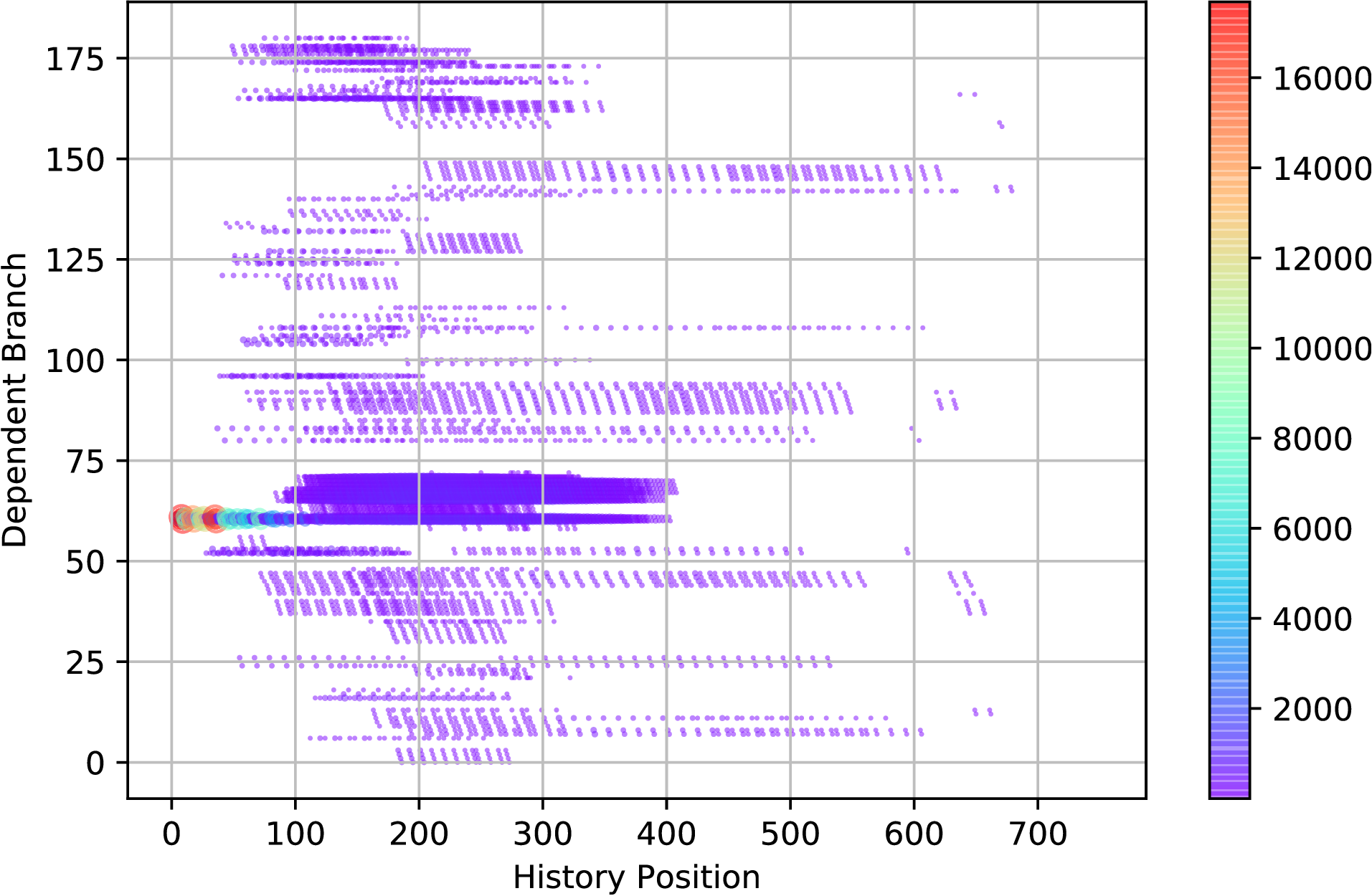}
    \label{fig:depbranch-omnetpp}
  }
  \subfloat[623.xalancbmk\_s]{
    \includegraphics[height=4cm]{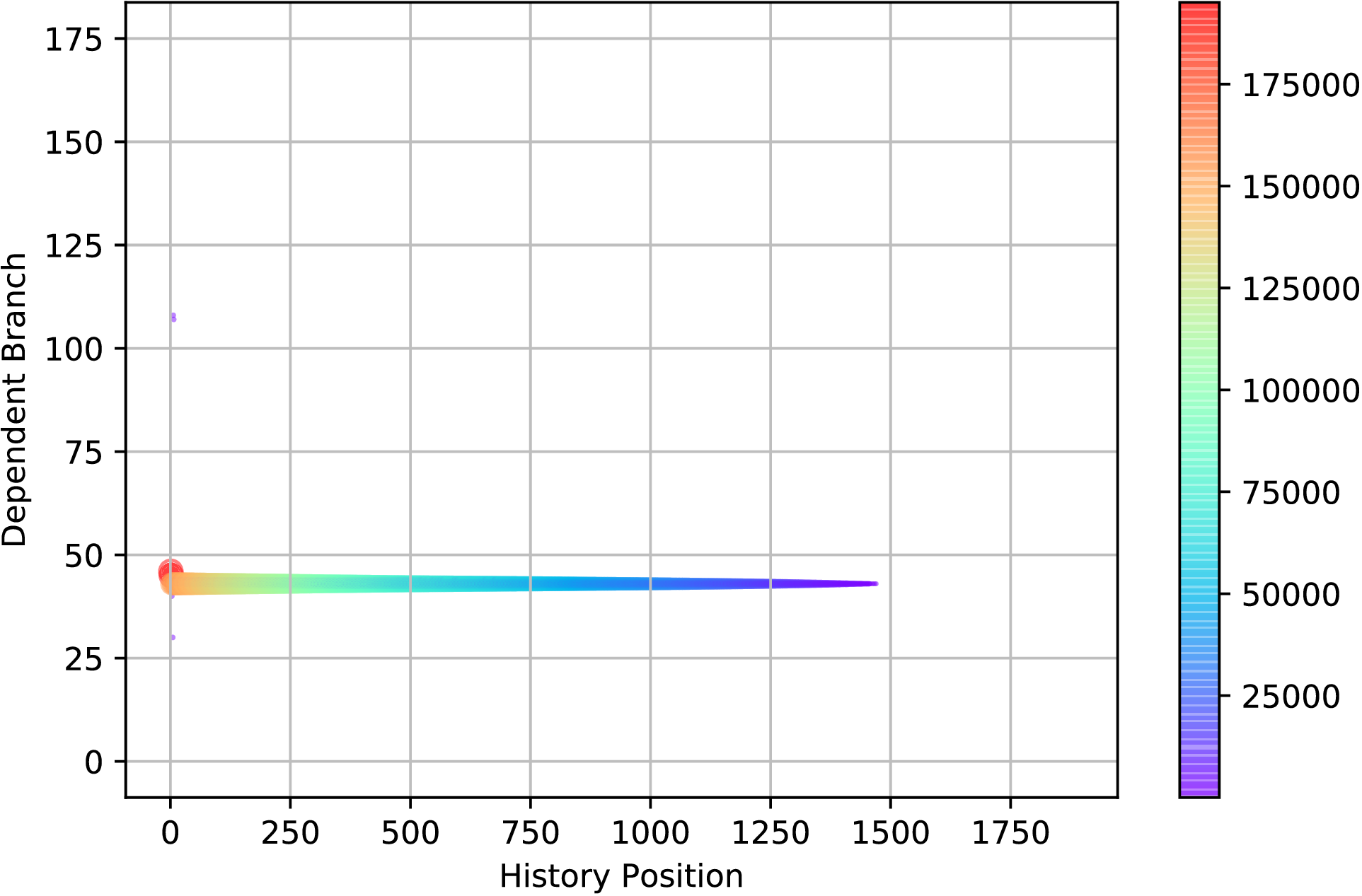}
    \label{fig:depbranch-xalancbmk}
  }
  \\
  \subfloat[625.x264\_s]{
    \includegraphics[height=4cm]{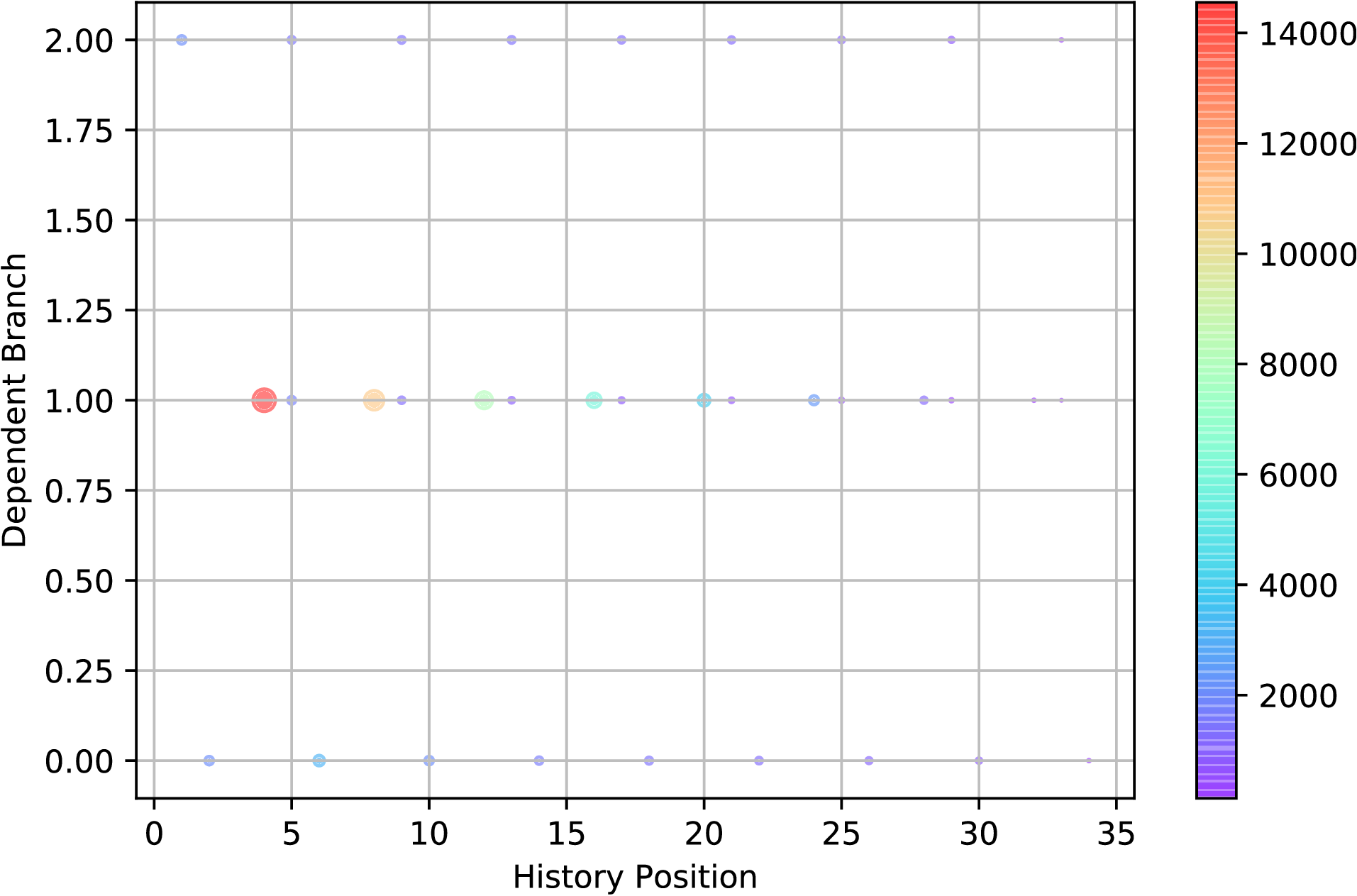}
    \label{fig:depbranch-x264}
  }
  \subfloat[631.deepsjeng\_s]{
    \includegraphics[height=4cm]{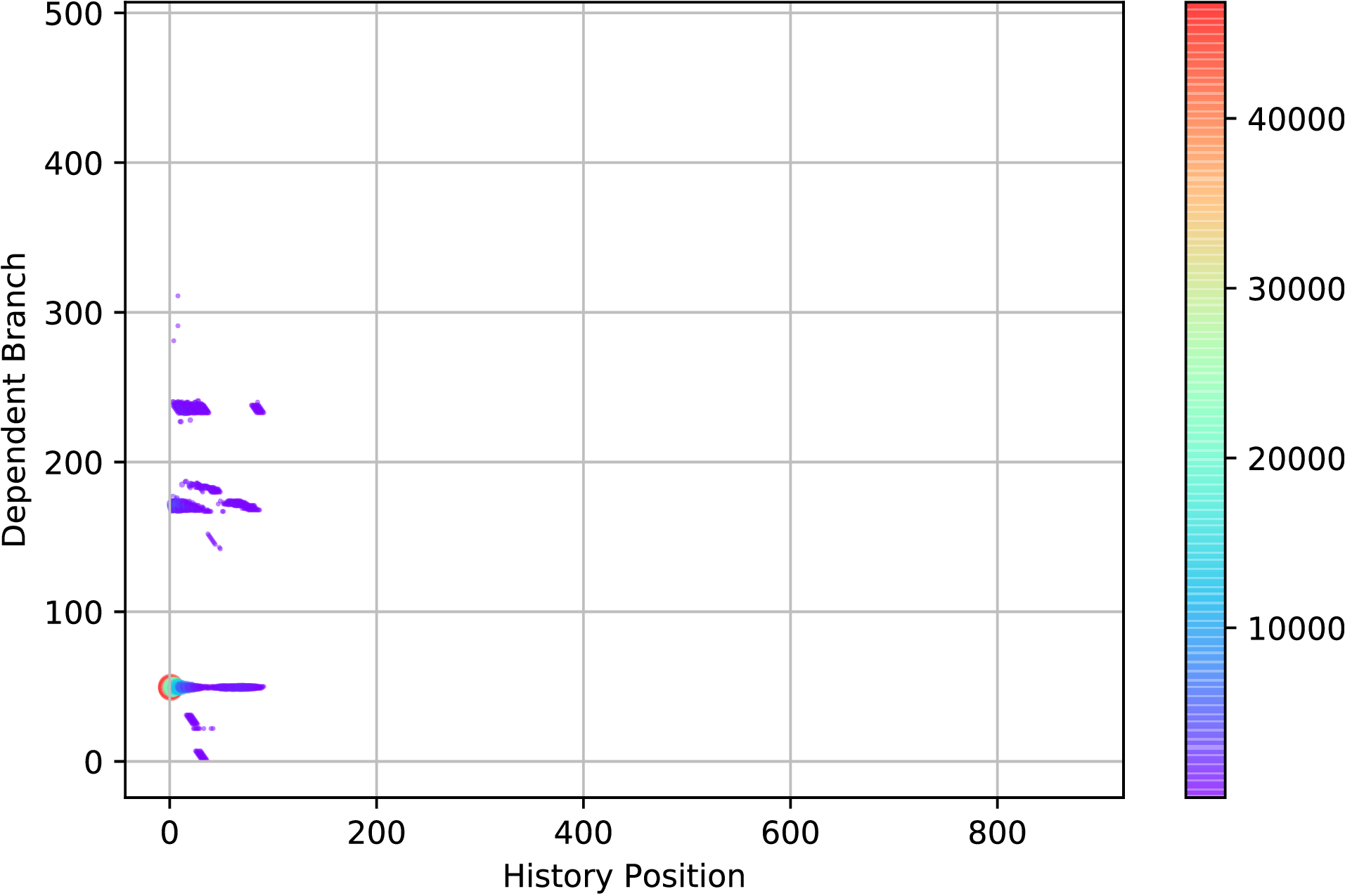}
    \label{fig:depbranch-deepsjeng}
  }
  \subfloat[641.leela\_s]{
    \includegraphics[height=4cm]{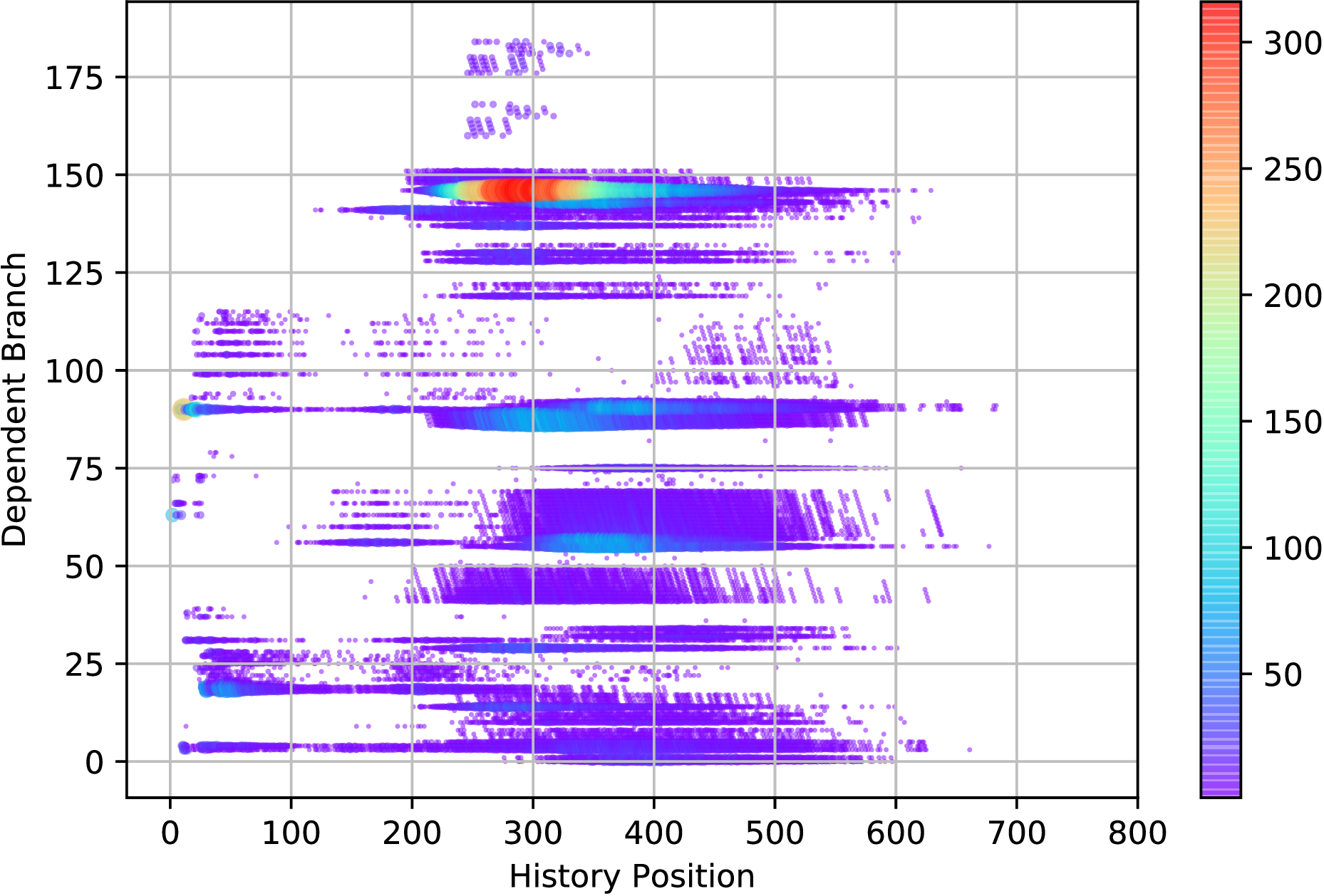}
    \label{fig:depbranch-leela}
  }
  \\
  \subfloat[641.exchange\_s]{
    \includegraphics[height=4cm]{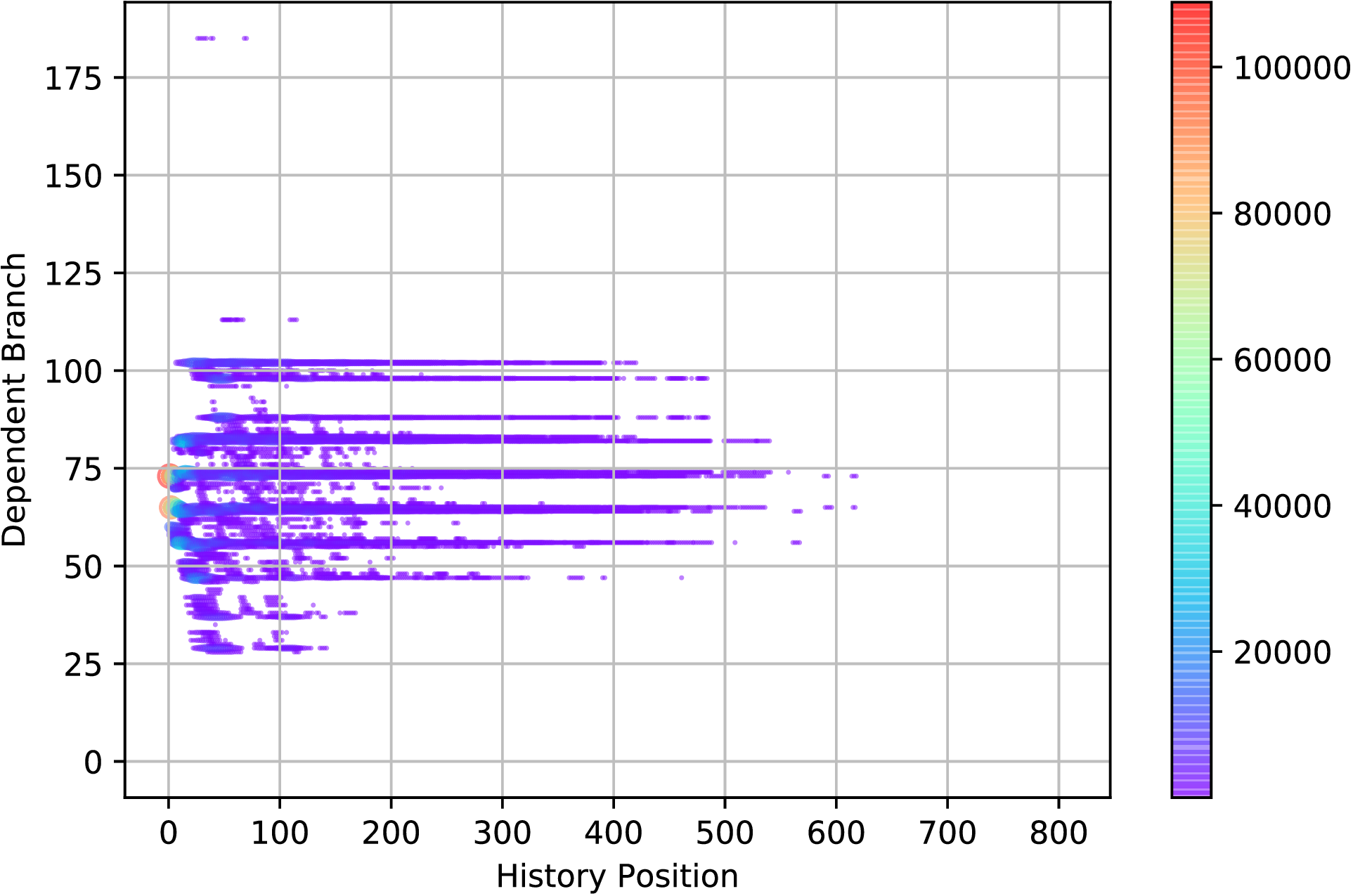}
    \label{fig:depbranch-exchange}
  }
  \subfloat[657.xz\_s]{
    \includegraphics[height=4cm]{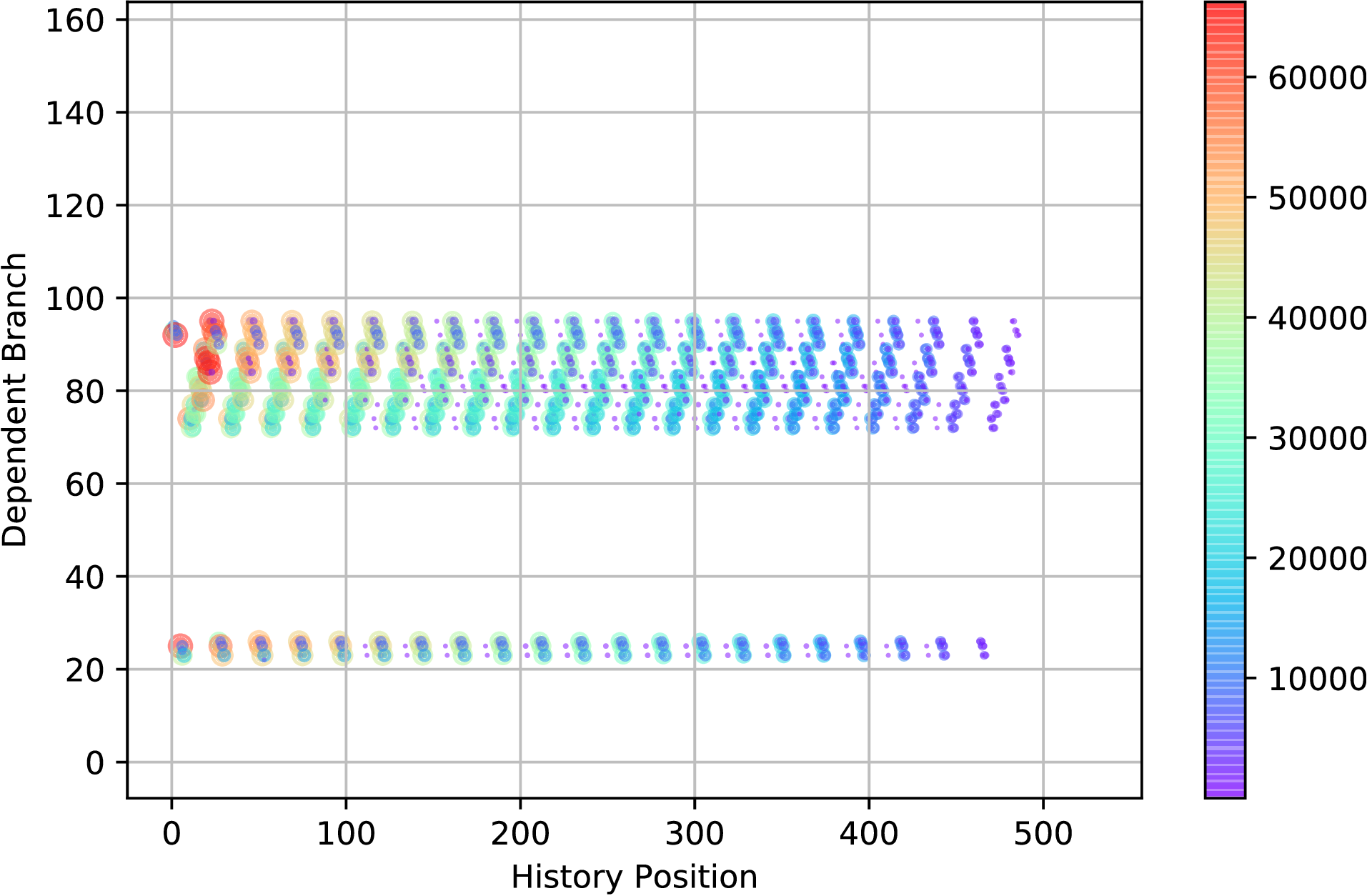}
    \label{fig:depbranch-xz}
  }
  \caption{Distribution of the history position of dependency branches
    for example H2P heavy hitters in each of the SPECint 2017
    benchmarks shown.  The size and color of each
    data point is proportional to the occurrence of that dependency
    branch at the corresponding history position.  Some data points
    are invisible because of low occurrence.}
  \label{fig:depbranches}
\end{figure*}

\subsection{H2Ps have high history variation}
\label{sec:h2p_high_var}
TAGE-SC-L 64KB tracks branch histories with lengths up to 3,000
(TAGE-SC-L 8KB allows up to 1,000).  However, we show that longer
history lengths inject \emph{more} variation into the predictive
signatures.  We characterize this variation by analyzing an H2P's
\emph{dependency branches}, i.e., previously-retired branches whose
conditions share an operand with the H2P, and are therefore predictive
at ground truth.  For a dynamic execution of an H2P, we compute
its operand dependency graph over the prior 5,000 instructions.  This
graph links instructions that read a common piece of data by tracking
chains of reads/writes to memory and registers.  We identify as a
dependency branch any prior conditional branch instruction that reads
a data value that is also read when computing the H2P's condition. For
each H2P, we perform this analysis for all of its dynamic executions
over the entire trace to produce a distribution over the history
positions of predictive dependency branches as they appear to the BPU.

Table~\ref{tab:dependency_extents} shows the min and max history
positions of these distributions and the number of dependency
branches for the top H2P heavy hitter of each SPECint 2017
benchmark.  We observe that the maximum history lengths across all
benchmarks fall within the history length limit of TAGE-SC-L 64KB.
This suggests that TAGE-SC-L 64KB has sufficient history to predict
the H2P, but that other factors contribute to its poor prediction
accuracy.  In Fig.~\ref{fig:depbranches}, we plot the distributions of
history positions for dependency branches associated with each heavy
hitter.  Notably, we see that any given dependency branch appears in
many different positions, and that the likelihood of it again
appearing in the same position is highly non-uniform. Together, this
data shows that predictions based on position-specific correlations or
the recurrence of exact patterns must contend with an enormous amount
of stochastic variation, and that variation increases with history
length.

\begin{table}[t!]
  \begin{tabularx}{\columnwidth}{l|r|r|r}
    \hline
    \hline
    Benchmark & Dep. Branches & Min Hist Pos & Max Hist Pos \\
    \hline
    \hline
    605.mcf\_s       &  43 & 2 & 1,221 \\
    620.omnetpp\_s   & 188 & 3 &   801 \\
    623.xalancbmk\_s & 176 & 1 & 1,879 \\
    625.x264\_s      &   3 & 1 &    34 \\
    631.deepsjeng\_s & 484 & 1 &   878 \\
    641.leela\_s     & 186 & 2 &   762 \\
    648.exchange2\_s & 167 & 1 &   863 \\
    657.xz\_s        & 157 & 1 &   530 \\
    \hline
    \hline
  \end{tabularx}
  \caption{Summary of dependency branch statistics for the top H2P heavy hitter
  branch in SPECint 2017 benchmarks.}
  \label{tab:dependency_extents}
\end{table}

We directly measure the effect of this variation by tracking how
TAGE-SC-L 64KB's table resources are allocated for H2P branches over
time.  TAGE-SC-L reserves entries for longer history lengths when the
longest-matching sequence produces mispredictions, while marking
incorrect or rarely used table entries for reallocation to other
branches.  Thus, H2P branches with high history sequence variation will
result in abnormally high reallocation rates.
Across our traces, we find that non-H2P branches are associated with
a small number of allocations and that their entries are rarely
reallocated---the median number of allocations per non-H2P branch is
4, while the median number of unique entries allocated to each is
also 4.  In contrast, H2P branches consume an outsized proportion of
table entries, with few of these entries producing useful predictions.
The median number of allocations per H2P is 13,093, while the median
number of unique table entries allocated to each H2P is only 3,990.
The discrepancy between these numbers is due to entries being
allocated, then scrapped for use by another branch, and eventually
being allocated for the same H2P branch once again.  On average, we
find that each non-H2P branch individually accounts for less than
0.01\% of allocations, whereas each H2P branch accounts for
3.6\%.  \emph{This behavior shows that TAGE-SC-L's underlying pattern matching
mechanism struggles to group predictive statistics in H2P history data, 
and that a large portion of storage resources are wasted as a result}.


\begin{figure*}[t!]
  \centering
  \subfloat[1x]{
    \includegraphics[clip,
      width=0.45\textwidth]{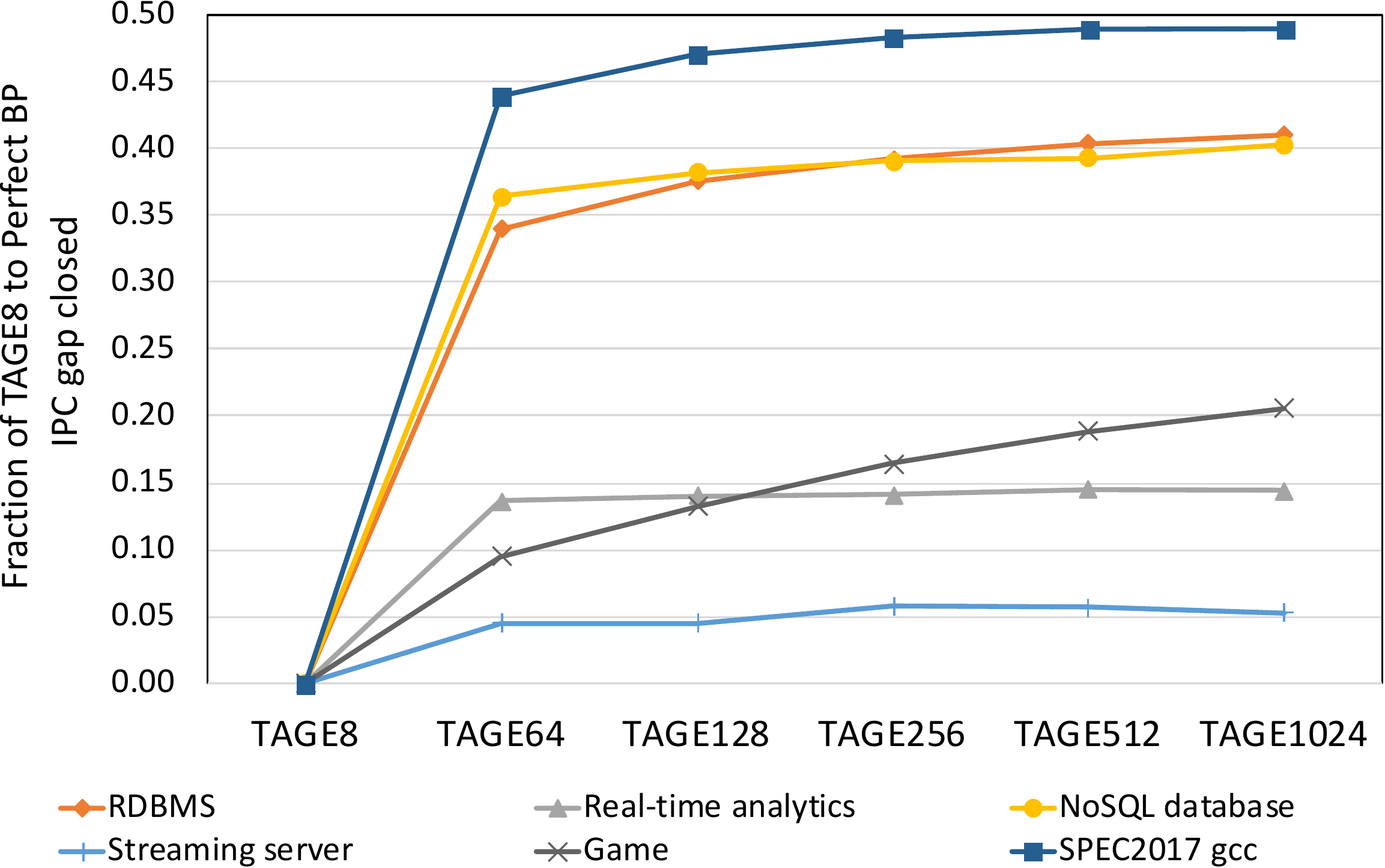}
    \label{fig:lcf_tage_scaling_1x}
  }\hfill
  \subfloat[2x]{
    \includegraphics[clip,
      width=0.45\textwidth]{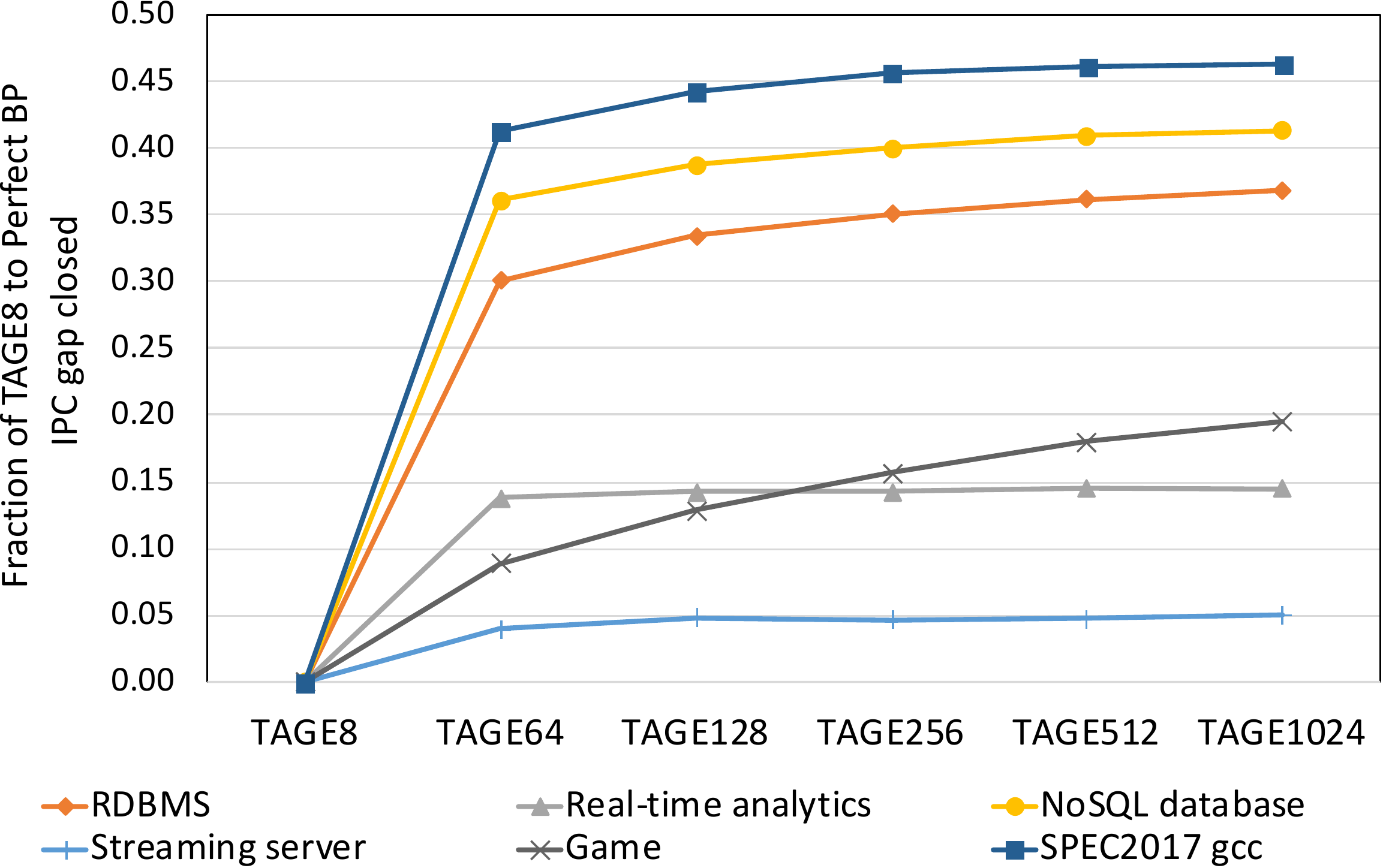}
    \label{fig:lcf_tage_scaling_2x}
  }
  \\
  \subfloat[4x]{
    \includegraphics[clip,
      width=0.45\textwidth]{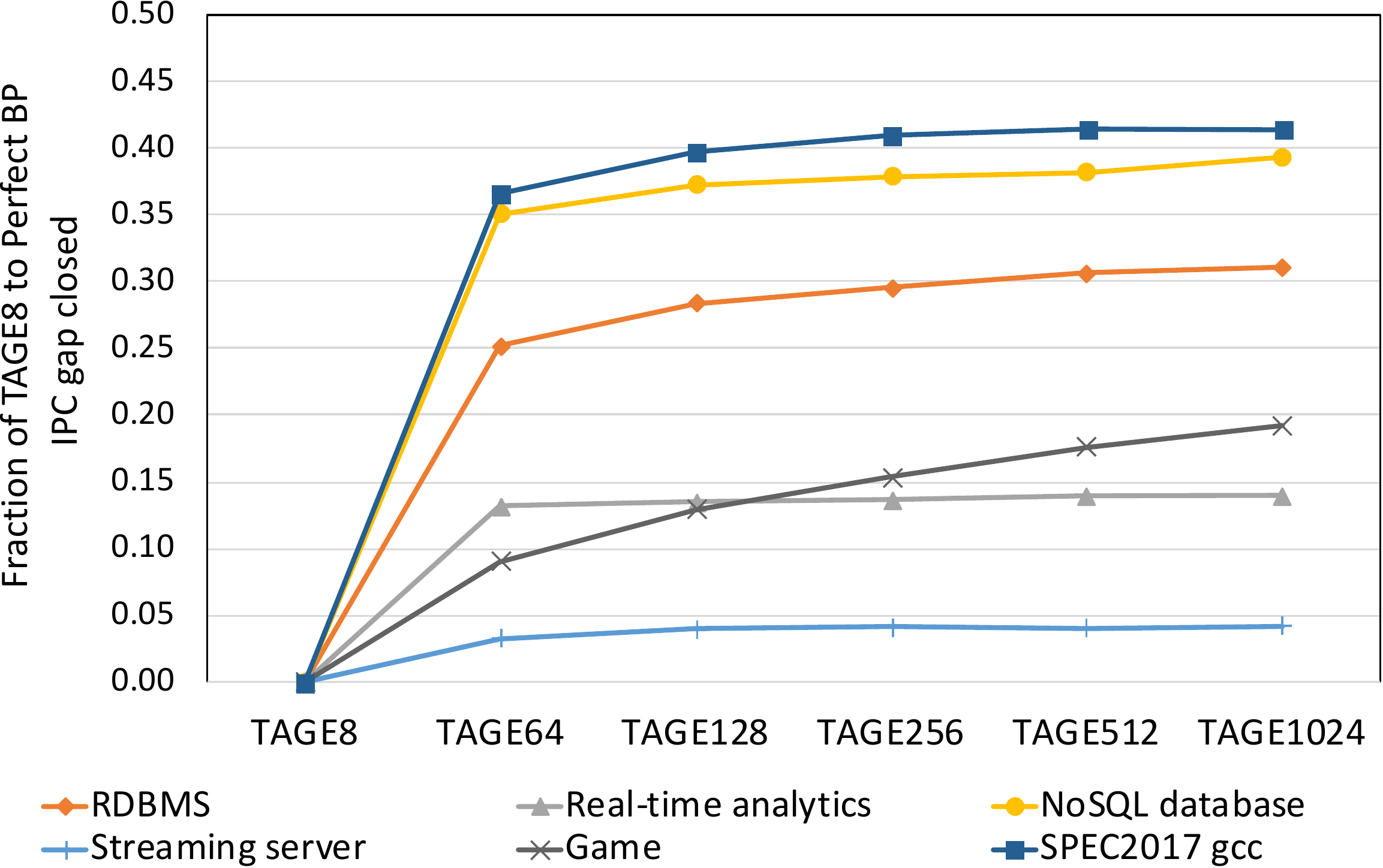}
    \label{fig:lcf_tage_scaling_4x}
  }\hfill
  \subfloat[8x]{
    \includegraphics[clip,
      width=0.45\textwidth]{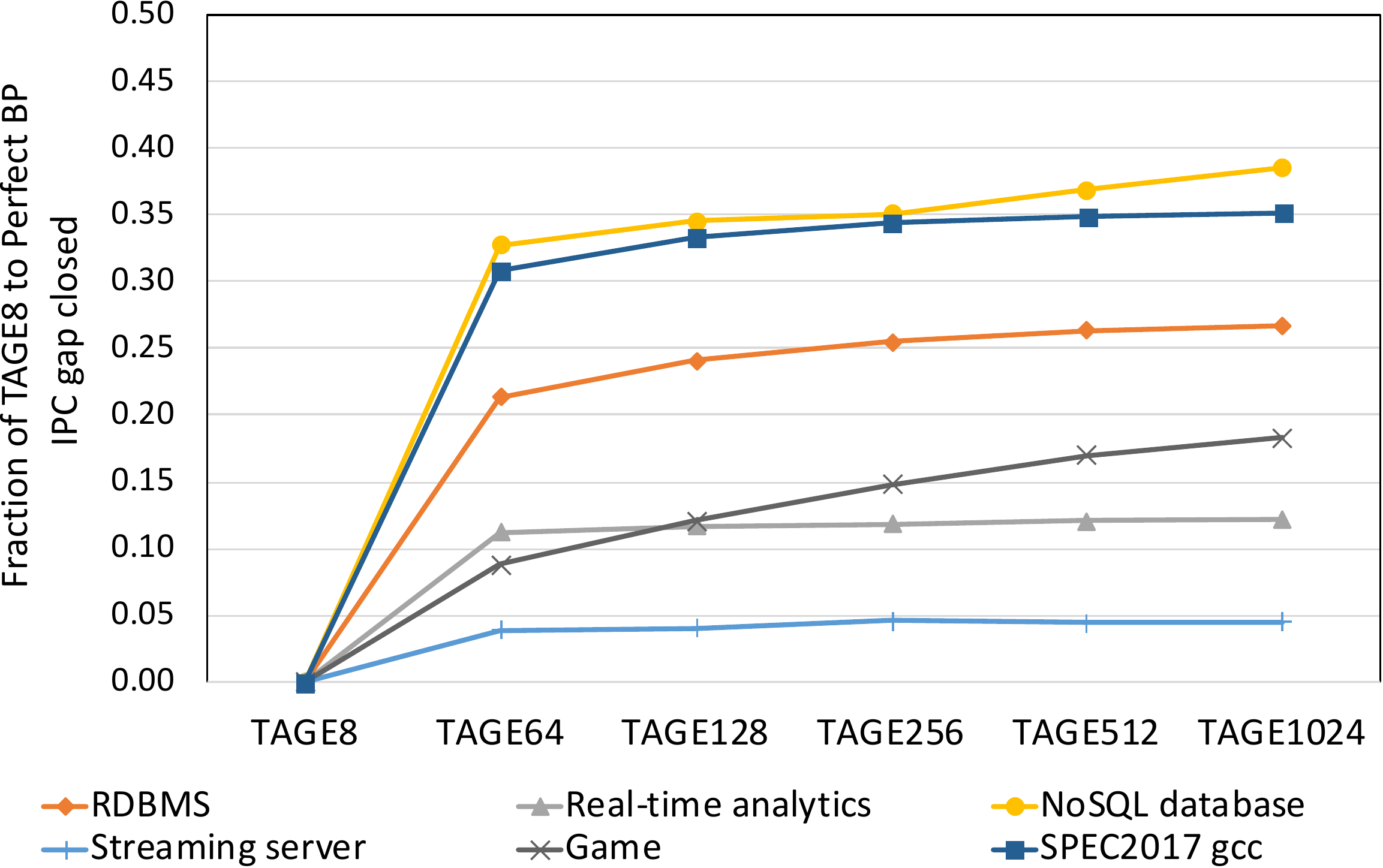}
    \label{fig:lcf_tage_scaling_8x}
  }
  \\
  \subfloat[16x]{
    \includegraphics[clip,
      width=0.45\textwidth]{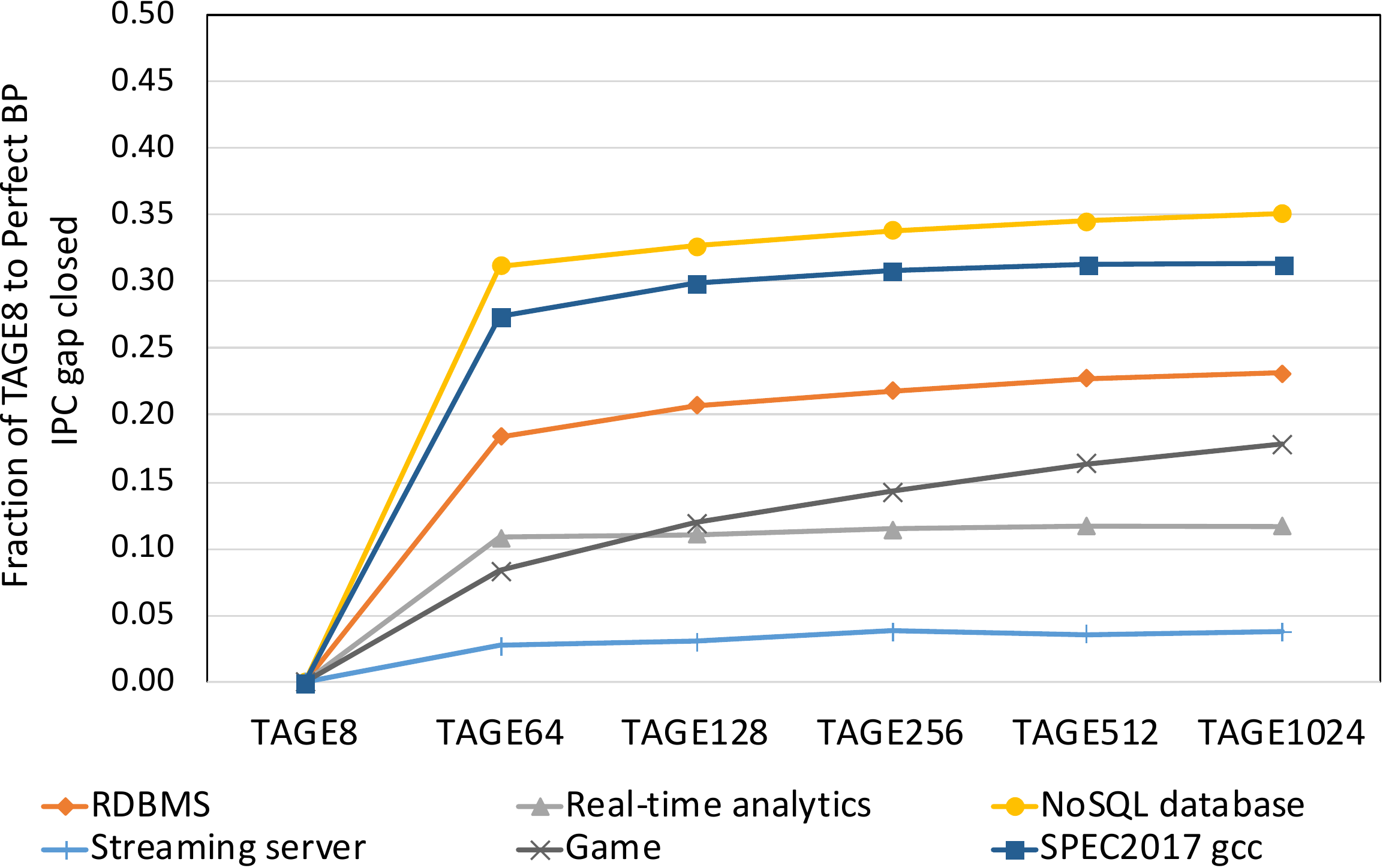}
    \label{fig:lcf_tage_scaling_16x}
  }\hfill
  \subfloat[32x]{
    \includegraphics[clip,
      width=0.45\textwidth]{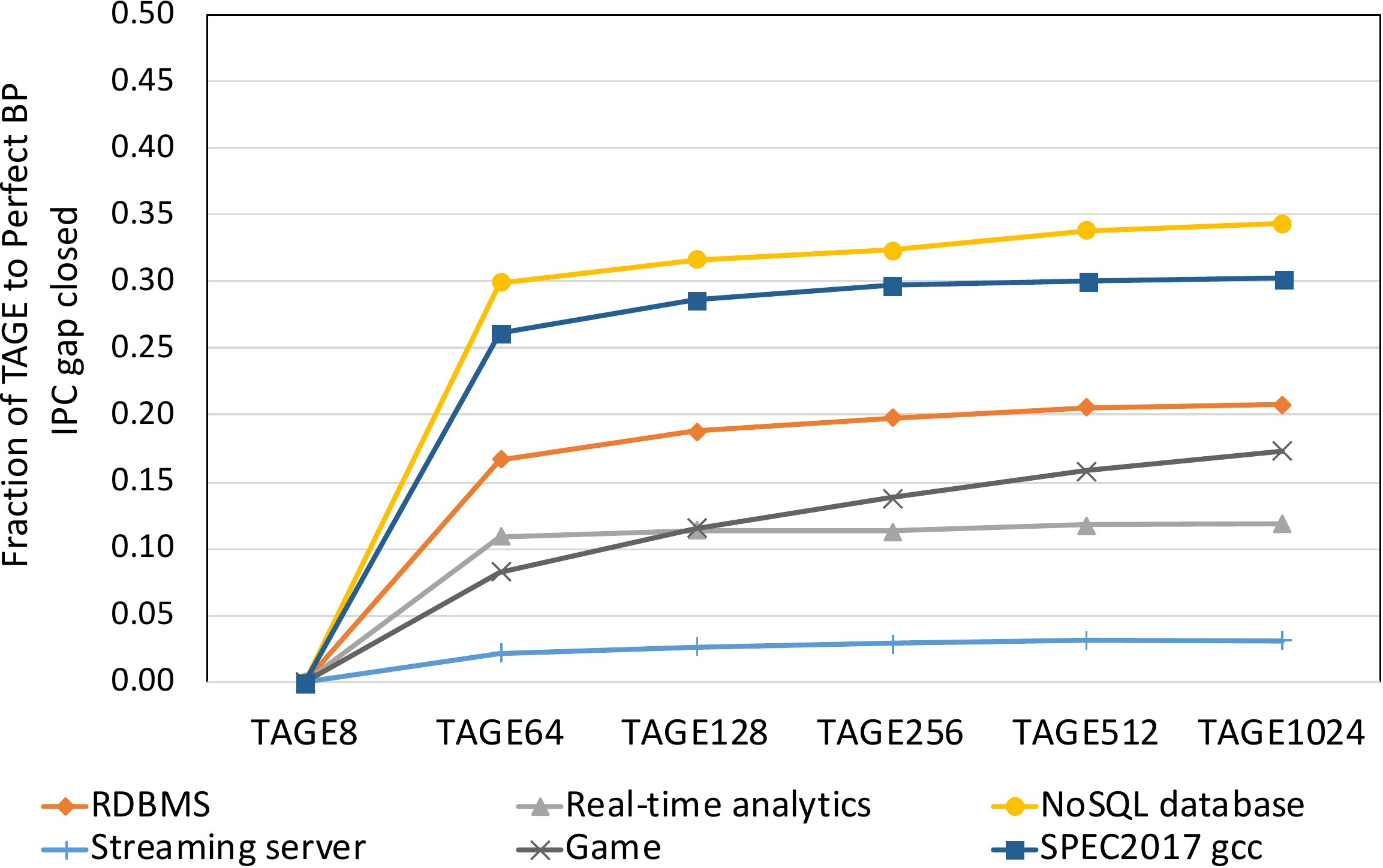}
    \label{fig:lcf_tage_scaling_32x}
  }  
  \caption{Scaling up CPU pipeline together with the number of table
    entries for TAGE-SC-L has diminishing returns.}
  \label{fig:lcf_tage_scaling}
\end{figure*}

\begin{figure}[ht!]
  \includegraphics[clip,
    width=\columnwidth]{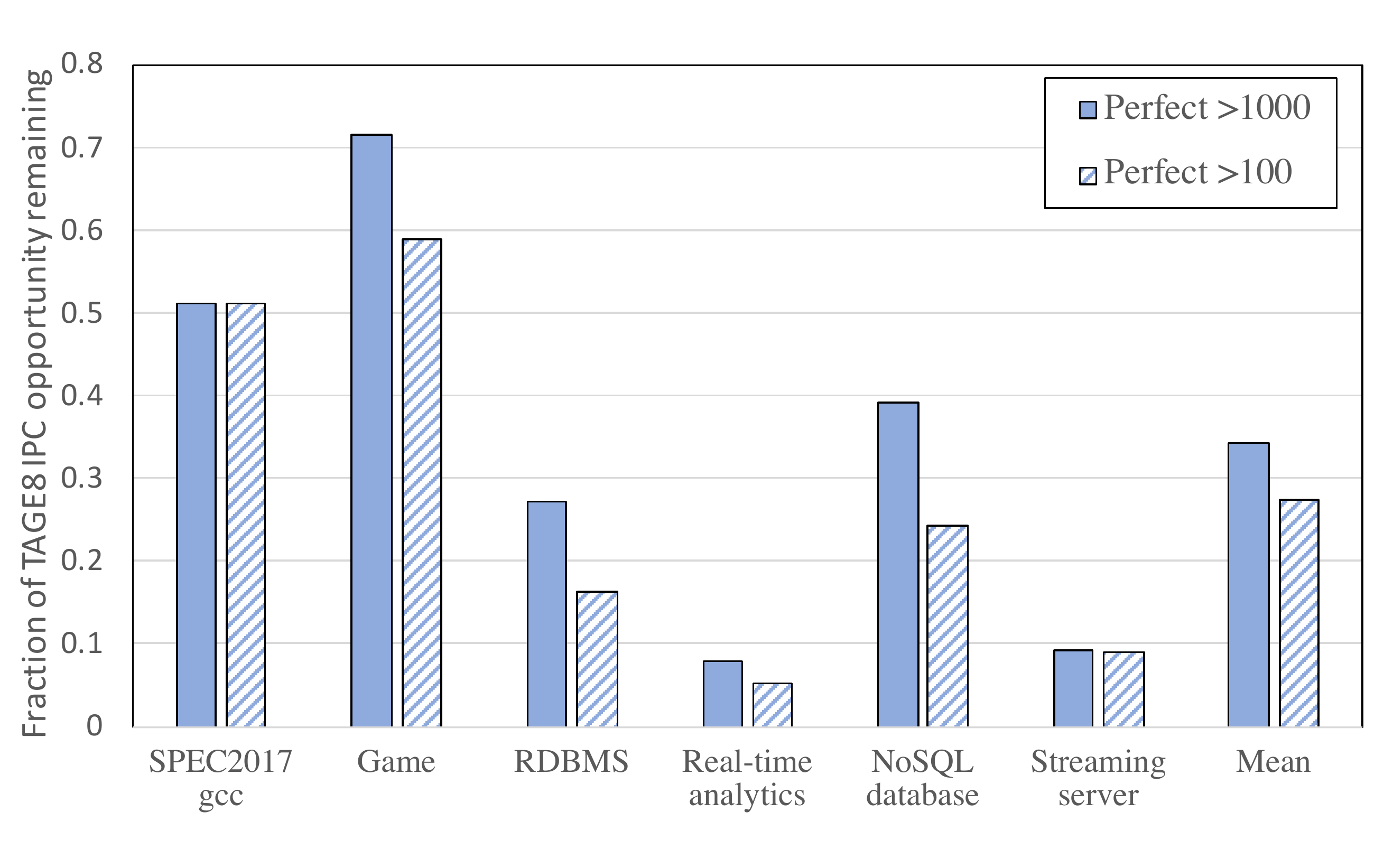}
  \caption{The fraction of IPC opportunity remaining even after perfectly
    predicting all branches with more than 1,000 (blue) and 100
    (orange) dynamic execution counts.}
  \label{fig:rare_branches_ipc}
\end{figure}

\subsection{Rare branches have poor statistics}
Table~\ref{tab:lcf_stats} indicates that LCF applications have a large
number of static branches that are only executed a handful
of times. This suggests that the baseline 8KB storage for TAGE-SC-L may 
quickly fill with entries that are not reused.  

In theory, simply increasing storage (i.e., increasing the table
capacity) to accommodate more of these branches should proportionately
improve IPC.  We perform a limit study that incrementally increases
total TAGE-SC-L storage from 8KB to 1024KB, as shown in
Fig.~\ref{fig:lcf_tage_scaling}.  For each LCF application, we define
the IPC opportunity as the IPC gap between TAGE-SC-L 8KB and perfect
branch prediction.  Then, at each storage size, we measure the portion
of the IPC opportunity captured by TAGE-SC-L.  We repeat this analysis
for different CPU pipeline scales to extrapolate to future designs.

It is immediately apparent that even at 1x (Skylake) pipeline scale,
TAGE-SC-L captures less than half of the IPC opportunity, even when afforded
an impractically high 1024KB of storage.  For almost all the measured
LCF applications, the greatest IPC gain comes from increasing storage
from 8KB to 64KB, after which improvements plateau.  Worse yet, as the
pipeline capacity is increased, scaling up storage yields dramatically
diminished returns---at 32x pipeline scale, a maximum of only 34\% of
the IPC opportunity is captured.

We show that this lack of further improvement despite additional resources
owes to that fact there are insufficient opportunities to both learn
and later reuse predictions for rare branches in LCF traces. 
In Section~\ref{sec:rare_branches}, Fig~\ref{fig:hist_lcf}
(middle) showed that, for LCF applications, 96\% of static branches
have fewer than 1,000 dynamic executions, 85\% have fewer than 100,
and that these rare branches have a wide spread in prediction accuracy
(Fig.~\ref{fig:lcf_acc_vs_dynexecs}). Using the largest 
TAGE-SC-L 1024KB storage configuration, we simulate 
the impact of predicting all branches with
more than 1,000 dynamic executions perfectly at 1x pipeline scale, and repeat 
this for all branches with more than 100 dynamic executions. These results
are reported in Fig.~\ref{fig:rare_branches_ipc}, and show that, on
average, 34.3\% of the IPC opportunity in large code footprint
application is due to rare branches (i.e., static branches with fewer
than 1,000 dynamic executions) and 27.4\% is due to the rarest
branches (i.e., static branches with fewer than 100 dynamic
executions). We observe that, \emph{with such a large portion of IPC
tied to branches with fewer than 100 dynamic executions over 30M instructions, 
rare branches supply too few statistics to support stable learning
\emph{and} later reuse at runtime.}

\section{New Directions for Branch Prediction}
\label{sec:newdirs}

The large IPC opportunity we have demonstrated and the inability of
existing branch predictors to tap into it suggests that the time is
ripe for reconsidering fundamental assumptions made in BPU design.
Since TAGE-SC-L already does so well on the vast majority of branches,
we argue that it should be left in place.  Any additional resources
given to the BPU should be devoted to augmenting it with other methods
that directly address the challenges laid out in Section~\ref{sec:scaling}.

\subsection{Reconsidering the deployment scenario}
\label{sec:deployment}

A key assumption in the design of branch predictors is the way in
which they are deployed. This can be categorized based on whether
predictive statistics are captured (i.e., trained) \emph{online} or
\emph{offline}, and similarly whether predictions are generated (i.e.,
via inference) \emph{online} or \emph{offline}.  Here, we define
\emph{online} as performing computations on the BPU and \emph{offline}
as employing computations that require data, such as branch history or
other microarchitectural state information, to be moved elsewhere.

Historically, branch predictors have assumed strict requirements that
training and inference could only be performed online.  However, this
scenario limits the power of the algorithms available for pattern
recognition, as well as their ability to exploit long-range
statistical relationships.  For example, the lightweight pattern
recognition mechanisms found in perceptron predictors make a best-case
assumption that pattern complexity is relatively low.  Meanwhile,
predictors such as TAGE-SC-L track statistics using low-bit width
saturating counters that allow for just a small, fixed amount of
hysteresis, and thereby make the implicit assumption that direction
statistics are stable only in the short-term.

To successfully resolve the issues that existing branch predictors
cannot, we argue that branch prediction techniques should not adhere
solely to online-training/online-inference assumptions.  By relaxing
the constraints to allow for \emph{offline training}, we open the door
not only to training over a much richer set of data, but also to
employing more powerful pattern recognition algorithms from machine
learning.

\subsection{Using richer training data}
\label{sec:addl_signals}

Branch predictors that solely perform online training are limited to
the data available within the BPU as an application runs. 
Constraints on BPU storage capacity, e.g. due to layout, further impose an
assumption that recent program state is the best predictive signal of
a branch outcome.  This ignores the possibility of contributions
coming from distant program states, such as prior executions of the
same SimPoint phase, or even statistics derived from prior application
executions.  Furthermore, online branch predictors must be application-agnostic, 
or general enough to perform universally well for all applications.

Offline training (for online prediction), however, has no such
limitations.  In particular, we argue that the key advantage to
offline training is the ability to train predictors
from a much larger set of statistics specific to a target application, 
for example aggregated over multiple executions. 
Successful offline training would thus rest upon \emph{collecting
  multiple long-duration traces of an application, executing over
  multiple distinct application inputs}, as in the trace collection
methodology we employed above. We note that this differs markedly from the relatively 
short single-input traces employed in CBP challenges, and corresponds to 
a large shift in the basic assumptions of current branch predictor development. 

For H2P branches, where long history lengths are useful for capturing
predictive signals in dependency branch correlations (recall
Fig.~\ref{fig:depbranches} in Section~\ref{sec:h2p_high_var}) but are
also the source of high history variation, offline training supports
the ability to identify ground-truth predictors such as dependency
branches.  One actionable way to exploit analyses similar to ours is
to design filters for a BPU to reduce data variation when predicting
H2Ps.  This would reduce the difficulty of on-BPU learning and make
more efficient use of limited on-BPU storage when folded into existing
algorithms such as TAGE-SC-L.

Offline training provides an additional method
to address LCF applications that we found to be capacity limited, i.e. 
those that saw performance gains when TAGE-SC-L
storage grew from 8KB to 64KB. This is because, under storage pressure at
8KB, TAGE must ``forget'' predictive patterns to make room for new ones. 
Identifying and storing
predictive signatures that are stable over the long term would reduce
the burden on TAGE to repeatedly relearn the same predictions. 

Offline training is particularly useful for the many pockets of code 
in LCF traces that execute infrequently over a single invocation of an application.
At the extreme, whenever an application is launched, TAGE learns from scratch,
no matter how many times the application had been invoked in the past.
This leaves on the table a significant opportunity for model reuse and
iterative refinement. Recording
statistics over multiple invocations of the same application---and
over distinct application inputs---increases the number and variation
in the samples of rare branches found in these pockets.  As a result,
offline training can yield more stable, higher confidence predictive
signatures for the rare branches that plague LCF traces than
online methods.

\begin{figure}[t!]
  \includegraphics[clip,width=\columnwidth]{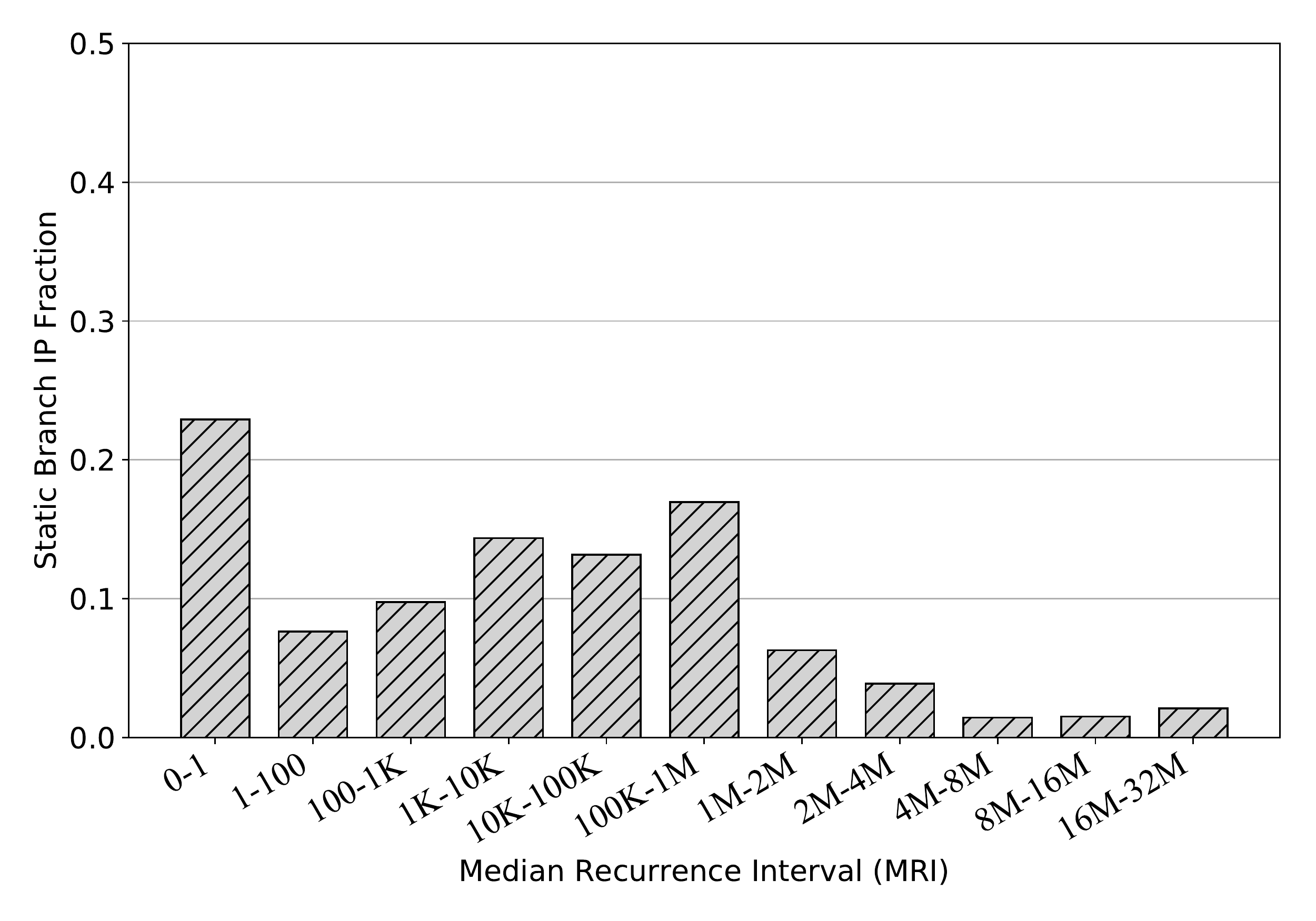}
  \caption{The distribution of the median recurrence interval of
    branches in the LCF dataset.  The recurrence interval is
    defined as the number of instructions between two consecutive
    dynamic executions of the same static branch IP.  The
    distribution indicates that phase-like behaviors on relatively
    long timescales exist in the LCF dataset, and that these phases can be
    exploited as an additional input signal to helper predictors. 
    }
  \label{fig:median_rref}
\end{figure}

Beyond generating a rich trace library, it is also possible to
incorporate data other than branch histories, from sources outside the
BPU.  One example is program phase information.  Program phases are a
well-known phenomenon~\cite{john2018} and can exist on different time
scales.  These can be inferred to a degree from the recurrence
interval of dynamic branches (i.e., the number of instructions between
two consecutive dynamic executions of the same static branch IP).  For
instance, if a branch has a very short recurrence interval, then it
may be part of a tight loop.  Conversely, a very large recurrence
interval may be indicative of a much more macro-level program phase,
spanning perhaps hundreds of thousands or even millions of basic
blocks.  Fig.~\ref{fig:median_rref} shows the distribution of the
median recurrence interval of static branch IPs in the LCF dataset.
In aggregate, the applications in this dataset have median recurrence
intervals peaking between 100,000 and 1,000,000 instructions (ignoring
the singleton branches in the first bin), suggesting that phases of
sufficient and varied sizes are available to exploit.  Phase
information can be derived from architectural counter values over
time, as in recent works~\cite{tarsaisca2019, tarsa2019prefetch}.
Conditioning branch histories on such phase information serves as
another way to improve training by reducing the variation in the
branch history data. 

\begin{figure*}[t!]
  \centering
  \subfloat[605.mcf\_s]{
    \includegraphics[clip,width=0.33\textwidth]{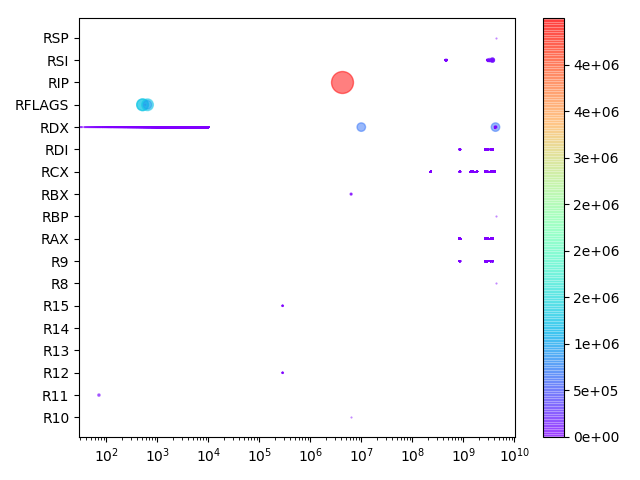}
    \label{fig:regvals_mcf}
  }
  \subfloat[620.omnetpp\_s]{
    \includegraphics[clip,width=0.33\textwidth]{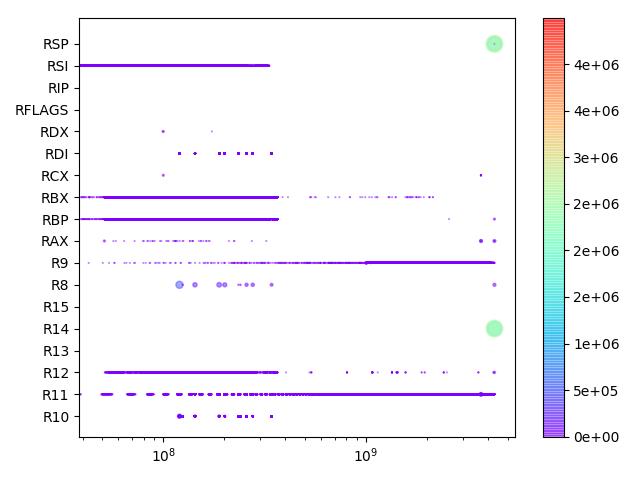}
    \label{fig:regvals_omnetpp}
  }
  \subfloat[625.x264\_s]{
    \includegraphics[clip,width=0.33\textwidth]{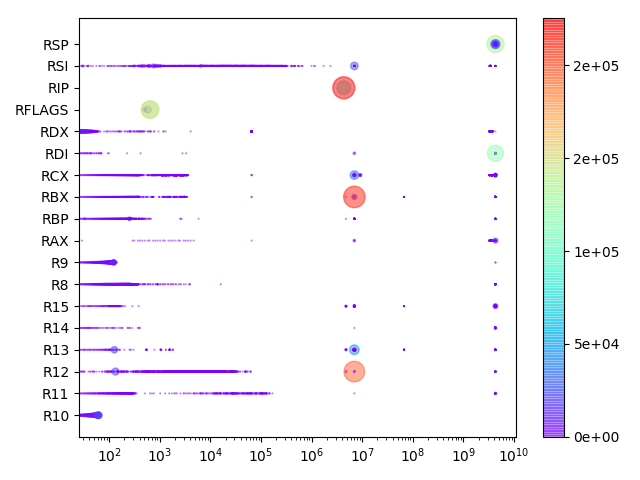}
    \label{fig:regvals_x264}
  }
  \\
  \subfloat[631.deepsjeng\_s]{
    \includegraphics[clip,width=0.33\textwidth]{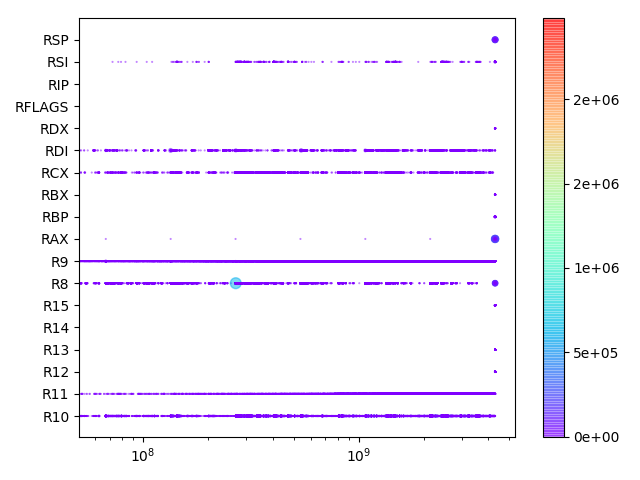}
    \label{fig:regvals_deepsjeng}
  }
  \subfloat[641.leela\_s]{
    \includegraphics[clip,width=0.33\textwidth]{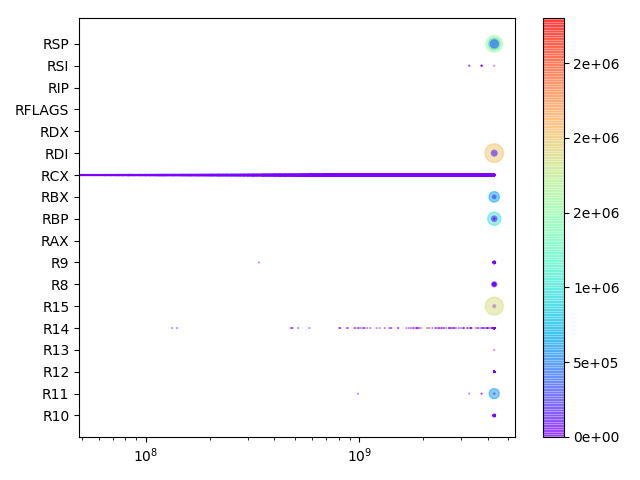}
    \label{fig:regvals_leela}
  }
  \subfloat[657.xz\_s]{
    \includegraphics[clip,width=0.33\textwidth]{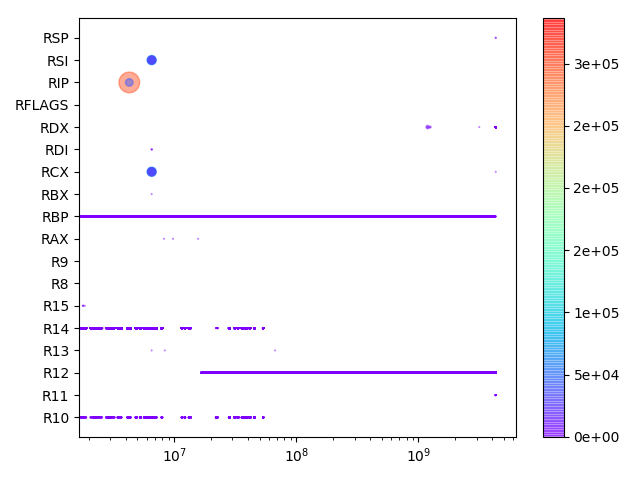}
    \label{fig:regvals_xz}
  }
  \caption{Distribution of register values written immediately preceding the
    top H2P heavy hitter branch in each of the SPECint2017 benchmarks.
    We record the bottom 32-bits of register writes in 18 tracked
    registers.  The x-axis plots the actual register value and is log
    scale. Each data point represents a register and a written value;
    its size and color is proportional to the number of times the
    value was written to the register.}
  \label{fig:regvals}
\end{figure*}

Yet another example of off-BPU information that could easily be
incorporated into offline training are the register values immediately
preceding a dynamic branch.  For data-dependent branches, which
typically resist prediction when using global branch histories alone,
this could be an additional correlative input signal to boost
prediction accuracy.  In Fig.~\ref{fig:regvals}, we plot the
distribution of the register values (lower 32-bits) written to each of
18 tracked registers, immediately preceding the dynamic executions of
the top H2P heavy-hitter in each of the SPECint 2017 benchmarks shown.
Two observations can be made readily: (1) that one distribution is
drastically different from the next, indicating that we should focus
on training branch-specific predictors; and (2) that there is complex
but recognizable structure in the distributions, suggesting that more
sophisticated machine learning algorithms such as neural networks may
be useful for extracting the underlying patterns.

\subsection{Using machine learning models}
By removing the requirement of online training, we also effectively
remove all computational constraints from the training process.
Whereas online training methods are limited to the storage and
operation complexity available on-BPU, training offline, e.g., over
the above trace datasets, can take advantage of the virtually
unlimited compute and storage resources of cloud computing
infrastructures.  This admits the use of powerful machine learning
algorithms such as convolutional neural networks (CNNs), which can
more fully extract patterns from the high-volume, high-complexity, and
high-variation H2P and rare branch data. We refer to powerful
predictors specialized to individual branches as \emph{helper
  predictors} since they are intended to be deployed alongside an
existing baseline predictor such as TAGE-SC-L.  In our companion paper
on CNN helper predictors~\cite{tarsa2019}, we show that \emph{models
  trained offline on applications traced over multiple inputs can
  generalize to unseen inputs}, thereby significantly improving online
prediction accuracy upon deployment.

Of course, performing inference online using models that were trained
offline can still be computationally expensive, but other
works~\cite{tarsaisca2019, beckmann2017maximizing, tarsa2019} have
shown that it is indeed feasible to implement even sophisticated
machine learning inference algorithms within current area and latency
constraints.  In the case of our CNN helper predictors, we take
advantage of low-precision (2-bit) neural
networks~\cite{courbariaux16} and a custom input encoding method to
simplify the forward pass (inference) computation to require just a
handful of bitwise operations.

\subsection{Amortizing offline training costs}
\label{sec:amortize}

One key issue that offline-training/online-inference predictors face
is whether the high cost of collecting comprehensive training data and
of offline training is worth the effort and resources.  We argue that
such an approach is particularly well-suited to data center
applications, where performance is of paramount importance and where
the training costs can be amortized through economies of scale.  Under
this framework, a customer's critical data center application would
first be instrumented and then traced. Subsequently, helper predictors
would be trained offline on the collected traces.  Once trained, the
predictors' model parameters (e.g., network weights in the case of a
CNN) could be stored as application metadata, e.g., under a new
segment type in an ELF binary.  The application would be installed on
machines across the data center, where each machine's OS would manage
loading the predictor(s) onto the BPU.  This ensures that gains in
prediction accuracy can be applied at scale.  Once this infrastructure
is in place, it then becomes possible to periodically update models
with additional training data.  In this way, helper predictors can be
iteratively refined over time, a key advantage over existing online
BPU mechanisms.  Of course, the exact mechanics of this deployment
model is the subject of future work, but we note that it is consistent
with other runtime hardware optimizations in the recent
literature~\cite{ravi2017charstar, tarsa2015machine,
  beckmann2017maximizing, tarsaisca2019}.

\section{Conclusions}
\label{sec:conc}

In this paper, we characterized branch mispredictions under the
state-of-the-art TAGE-SC-L branch predictor.  Using SPECint 2017
benchmarks and a set of large code footprint applications, we
demonstrate that there remains an untapped IPC opportunity due to
these mispredictions, the size of which is on par with advancing
process technology.  We identified hard-to-predict (H2P) and rare
branches as two classes of branches whose mispredictions account for
this missed IPC opportunity, and showed that simply scaling up the
storage capacity of TAGE-SC-L global history tables does not rescue
these mispredictions.

From these measurements and analyses, it is clear that branch
prediction is far from being solved and that there remains substantial
headroom for BPU improvement.  In response, we propose new assumptions
for branch predictor development---namely, \emph{offline training} on
data aggregated over multiple application executions.  This approach
enables new research directions that directly address the causes of
mispredictions for both H2P branches and rare branches.  These include
exploiting more diverse training data to improve statistical power for
rare branches, as well as additional computational resources to train
specialized helper predictors for specific branches.

%
%

\bibliographystyle{ieeetr}
\bibliography{ref}
\end{document}